\documentclass[preprint]{emulateapj}
\slugcomment{{\sc Accepted to ApJ:} May 20, 2013} 
\usepackage{epstopdf}
\usepackage{amsmath,amsopn,amsxtra,txfonts}
\usepackage{comment} 
\usepackage{natbib} 
\usepackage{graphicx}

\newcommand{\ha}{H$\alpha$}

\newcommand{\kms}{ \ifmmode{\rm km\thinspace s^{-1}}\else km\thinspace s$^{-1}$\fi}

\newcommand{\pc}{\ensuremath{ \, \mathrm{pc}}}
\newcommand{\kpc}{\ensuremath{\, \mathrm{kpc}}}
\newcommand{\cm}{\ensuremath{ \, \mathrm{cm}}}

\newcommand{\s}{\ensuremath{ \, \mathrm{s}}}

\newcommand{\sr}{\ensuremath{ \, \mathrm{sr}}}

\newcommand{\dg}{\ifmmode{^{\circ}}\else $^{\circ}$\fi}

\newcommand{\hi}{\ion{H}{1}}

\newcommand{\lb}{\ifmmode{(\ell,b)} \else $(\ell,b)$\fi}
\newcommand{\nii}{[\ion{N}{2}]}

\newcommand{\sii}{[{\ion{S}{2}}]}

\newcommand{\oi}{[\ion{O}{1}]}

\newcommand{\vlsr}{\ifmmode{v_{\rm{LSR}}}\else $v_{\rm{LSR}}$\fi}
\newcommand{\av}{\ifmmode{A(V)}\else $A(V)$\fi}
\newcommand{\ebv}{\ifmmode{E(B-V)}\else $E(B-V)$\fi}
\newcommand{\iha}{\ifmmode{I_{\rm{H}\alpha}} \else $I_{\rm H \alpha}$\fi}
\newcommand{\ihb}{\ifmmode{I_{\rm{H}\beta}} \else $I_{\rm H \beta}$\fi}
\newcommand{\isii}{\ifmmode{I_{\ion{\rm{S}}{2}}} \else $I_{\rm [S \textsc{ ii}]}$\fi}
\newcommand{\inii}{\ifmmode{I_{\ion{\rm{n}}{2}}} \else $I_{\rm [N \textsc{ ii}]}$\fi}
\newcommand{\ioi}{\ifmmode{I_{\ion{\rm{o}}{1}}} \else $I_{\rm [O \textsc{ i}]}$\fi}

\newcommand{\Hplus}{\ensuremath{\mathrm{H}^+}}

\newcommand{\mha}{\ensuremath{\mathrm{H} \alpha}}

\newcommand{\mnii}{\ensuremath{\textrm{[\ion{N}{2}]}}}
\newcommand{\msii}{\ensuremath{\textrm{[{\ion{S}{2}}]}}}

\newcommand{\N}{\ensuremath{\mathrm{N}}}
\renewcommand{\H}{\ensuremath{\mathrm{H}}}

\newcommand{\vhi}{$\bar{v}_{\rm{H}\textsc{ i}}$}

\newcommand{\nhi}{$N_{\rm{H}\textsc{ i}}$}

\newcommand{\lhi}{$L_{\rm{H}\textsc{ i}}$}

\newcommand{\vgeo}{\ifmmode{v_{\mathrm{geo}}} \else $v_{\mathrm{geo}}$\fi}

\shorttitle{Gas Inflow Associated with High-Velocity Cloud Complex~A}
\shortauthors{Barger et al.}

\begin{document}
%\nocite{*} This will cause ALL references in Reference.bib to be listed in the Bibliography 

\author{K. A. Barger\altaffilmark{1}, L. M. Haffner, B. P. Wakker, Alex. S. Hill\altaffilmark{2}}
\affil{Department of Astronomy, University of Wisconsin-Madison, Madison, WI 53706, USA}
\email{kbargers@nd.edu}
\email{haffner@astro.wisc.edu}
\email{Alex.Hill@csiro.au}
\email{wakker@astro.wisc.edu}

\altaffiltext{1}{Now at the Department of Physics, University of Notre Dame, South Bend, IN 46556, USA}
\altaffiltext{2}{Now at the CSIRO Astronomy \& Space Science, P.O. Box 76, Epping NSW 1710, Australia}

\author{G. J. Madsen}
\affil{Sydney Institute for Astronomy, School of Physics, University of Sydney, NSW 2006, Australia}
\email{greg.madsen@sydney.edu.au}

\and

\author{A. K. Duncan}
\affil{Rose-Hulman Institute of Technology, Terre Haute, IN 47803, USA}

%\email{kbarger@astro.wisc.edu, hill@astro.wisc.edu, haffner@astro.wisc.edu, wakker@astro.wisc.edu, madsen@aao.gov.au}

\title{Present-day Galactic Evolution: Low-metallicity, Warm, Ionized Gas Inflow Associated with High-Velocity Cloud Complex~A}

\begin{abstract}

The high-velocity cloud (HVC) Complex~A is a probe of the physical conditions in the Galactic halo. The kinematics, morphology, distance, and metallicity of Complex A indicate that it represents new material that is accreting onto the Galaxy. We present Wisconsin \ha\ Mapper (WHAM) kinematically resolved observations of Complex~A over the velocity range of $-250$ to $-50~\kms$ in the local standard of rest reference frame. These observations include the first full \ha\ intensity map of Complex~A across $(\mathit{l, b}) = (124\arcdeg, 18\arcdeg)$ to $(171\arcdeg, 53\arcdeg)$ and deep targeted observations in \ha, \sii $\lambda6716$, \nii $\lambda6584$, and \oi $\lambda6300$ towards regions with high \hi\ column densities, background quasars, and stars. The \ha\ data imply that the masses of neutral and ionized material in the cloud are similar, both being greater than $10^6\, M_{\odot}$. We find that the \citet{1999ApJ...510L..33B, 2001ApJ...550L.231B} model for the intensity of the ionizing radiation near the Milky Way is consistent with the known distance of the high-latitude part of Complex~A and an assumed cloud geometry that puts the lower-latitude parts of the cloud at a distance of 7 to 8~\kpc. This compatibility implies a 5\% ionizing photon escape fraction from the Galactic disk. We also provide the nitrogen and sulfur upper abundance solutions for a series of temperatures, metallicities, and cloud configurations for purely photoionized gas; these solutions are consistent with the sub-solar abundances found by previous studies, especially for temperatures above $10^4~{\rm K}$ or for gas with a high fraction of singly-ionized nitrogen and sulfur. 

\end{abstract}


\maketitle

\section{Introduction}

Low-metallicity inflows of gas influence galactic evolution by providing new material for star formation and thereby modifying the metallicity of the interstellar medium and the stars within galaxies. Many high-velocity clouds (HVCs) show indications of interaction with the Milky Way including Complex~A, the Smith Cloud (Complex GCP), and the Magellanic Stream (e.g., \citealt{2000A&A...357..120B, 2008ApJ...679L..21L, 2009IAUS..254..241B}). Other nearby galaxies, such as M31, also seem to have HVCs interacting with them (e.g., \citealt{2005ASPC..331..113T}, \citealt{2005ASPC..331..105W}). These clouds exhibit velocities that are inconsistent with Galactic rotation, generally defined as a local standard of rest (LSR) velocity greater than {$90$ km s$^{-1}$}. Accreted HVCs can dilute the gas in galactic systems and modify their chemical evolution. Understanding the metallicity history of the Milky Way, partly revealed through the G-dwarf problem (\citealt{1962AJ.....67..486V, 1963ApJ...137..758S, 1975MNRAS.172...13P}), and its star formation rate requires understanding these clouds.

The origin of most of these HVCs remains a mystery. No universal mechanism exists to explain their formation, so the origin of each cloud requires investigation. Considerable work has been done to uncover these origins. \citet{1966BAN....18..421O, 1970A&A.....7..381O}  already proposed most of the mechanisms
that are still under consideration today, but at present, it is possible to point to specific examples for three of the processes that he suggested. The Magellanic System is the nearest example of a galaxy interaction producing high-velocity gaseous structures with material strewn throughout the halos of the Milky Way and the Magellanic Clouds. Models show that tidal forces and ram pressure can partly---if not dominantly---produce this stripped material as the clouds encounter each other and the Milky Way (e.g., \citealt{1977MNRAS.181...59L, 1980PASJ...32..581M, 1982MNRAS.198..707L, 1985PASAu...6..195M, 1996MNRAS.278..191G, 1994MNRAS.270..209M} cf., \citealt{2010ApJ...721L..97B}). Other than galaxy interactions, plausible sources of HVCs include residual primordial gas from the formation of the universe and material returning as part of a galactic fountain \citep{1980ApJ...236..577B}. 

\begin{figure*}
\begin{center}
\includegraphics[scale=.60,angle=0]{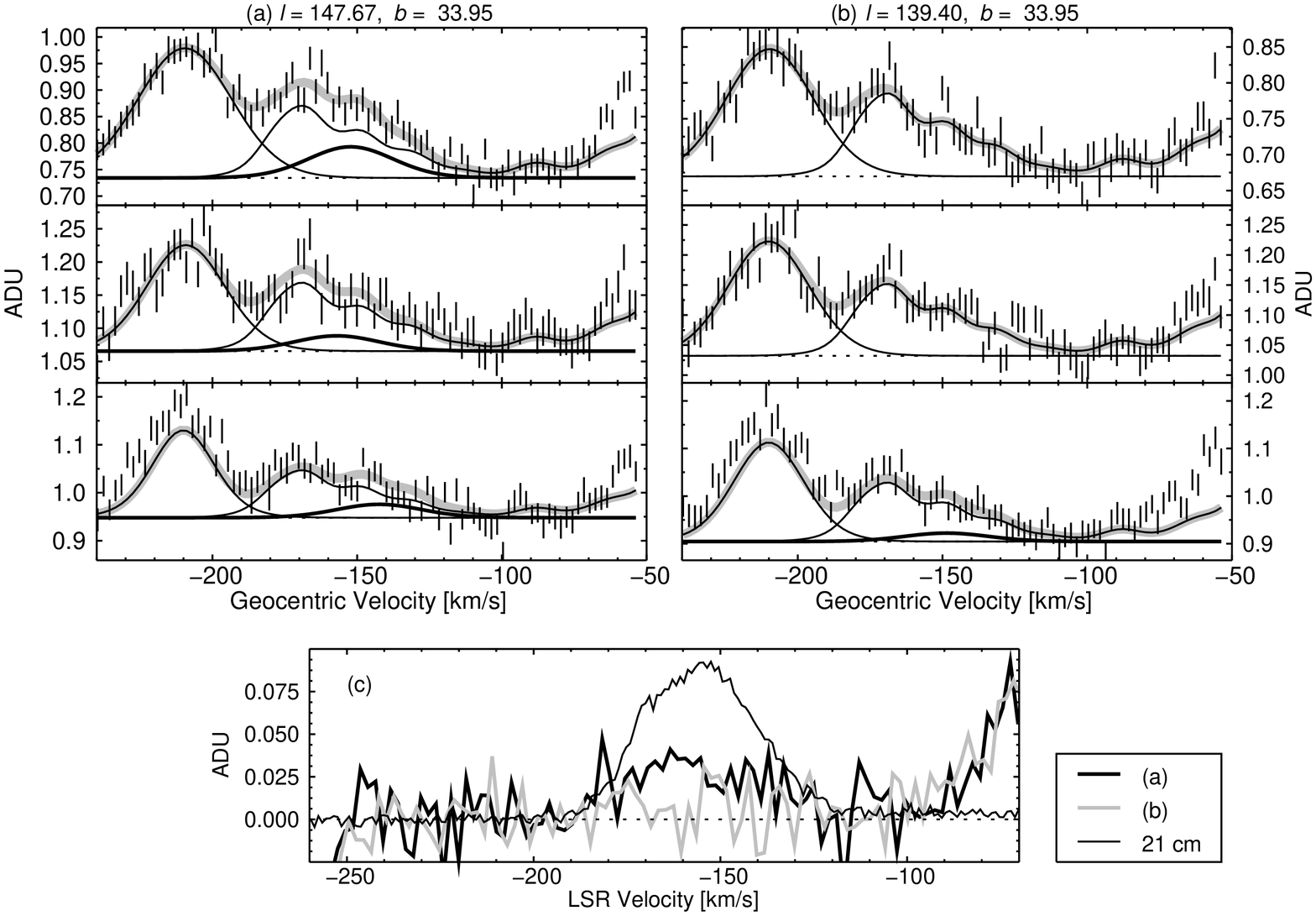}
\end{center}
\figcaption{Reduction of WHAM spectra toward two directions. Panels \emph{(a)} and \emph{(b)} display three 60-second exposures taken over three nights: 2005 Feb (bottom), 2005 Mar (middle), and 2006 Apr (top). Panel \emph{(a)} is near sightline core~AIII while \emph{(b)} is a sightline about 7\arcdeg\ away in a region with no detected 21~cm emission. Panels \emph{(a)} and \emph{(b)} show unreduced data. The dotted lines denote the fitted baselines for each spectrum. Overlaid in each panel is the full fit constructed to reduce the data and extract Galactic emission. The thin solid line with multiple components is the composite atmospheric template of faint emission described in Section \ref{sec:mapping}. The bright Gaussian component traced by a thin solid line at $\sim-210$\kms\ represents a geocoronal ``ghost'' feature. The thick black line peaking near -150\kms\ marks the \ha\ emission from Complex~A. The thick gray line is the resulting total fit that models the data. After subtracting the baseline, atmospheric template, and the ``ghost'' Gaussian component, the three spectra are translated to the LSR velocity frame and averaged. The bottom panel shows the reduced spectra for these two directions. As noted in the text, the rise at the red edge of the spectra is due to lower-velocity Galactic emission along these lines-of-sight. The intensity is given in arbitrary data units, where $1~{\rm ADU}\sim22.8~{\rm R}~(\kms)^{-1}$.
\label{figure:MapSpectraSample}}
\end{figure*}

%Please make these figures as large as possible. Please use an entire page to display them. 
\begin{figure*}
\begin{center}
\includegraphics[scale=.35,angle=0]{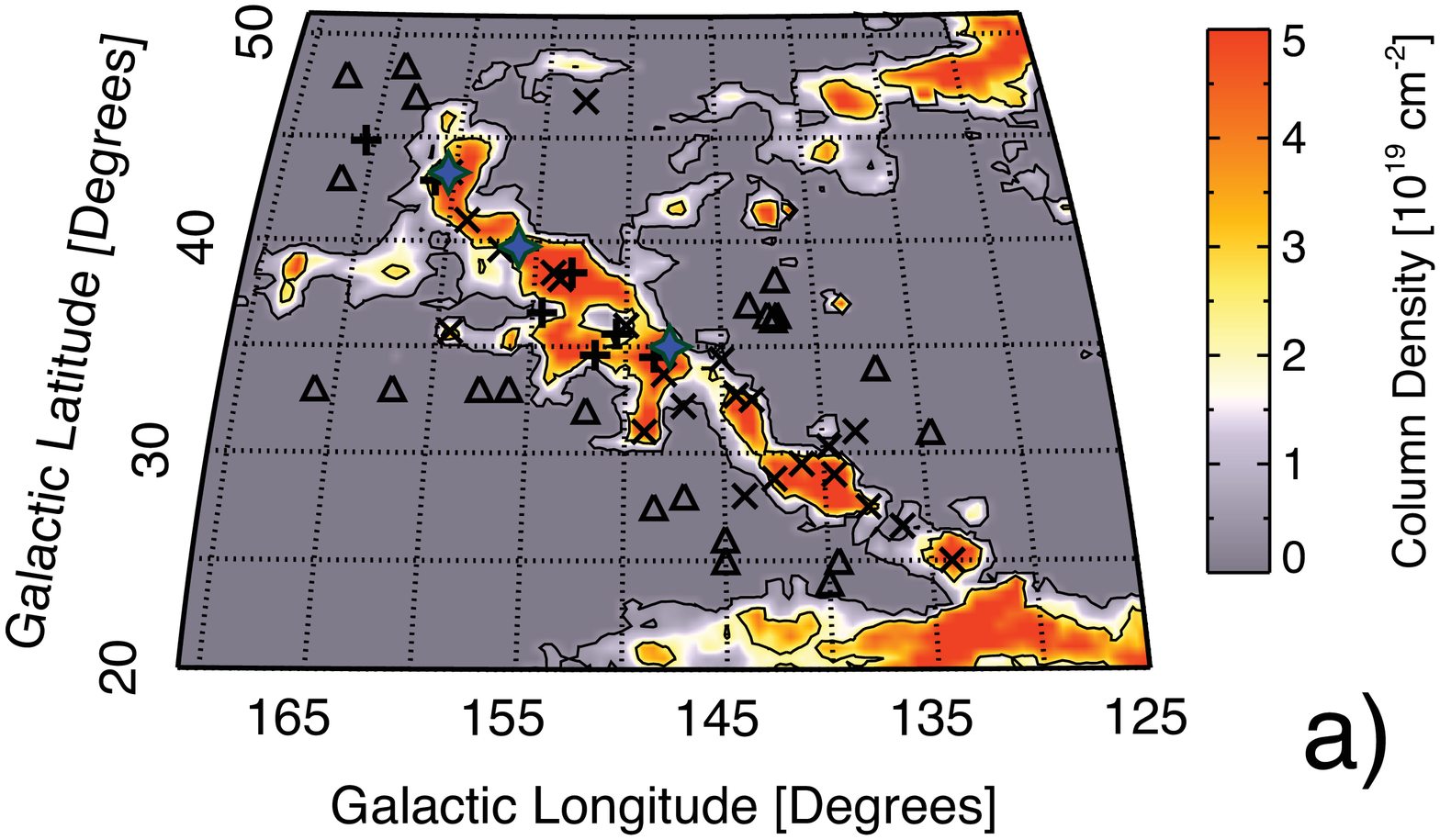}\includegraphics[scale=.35,angle=0]{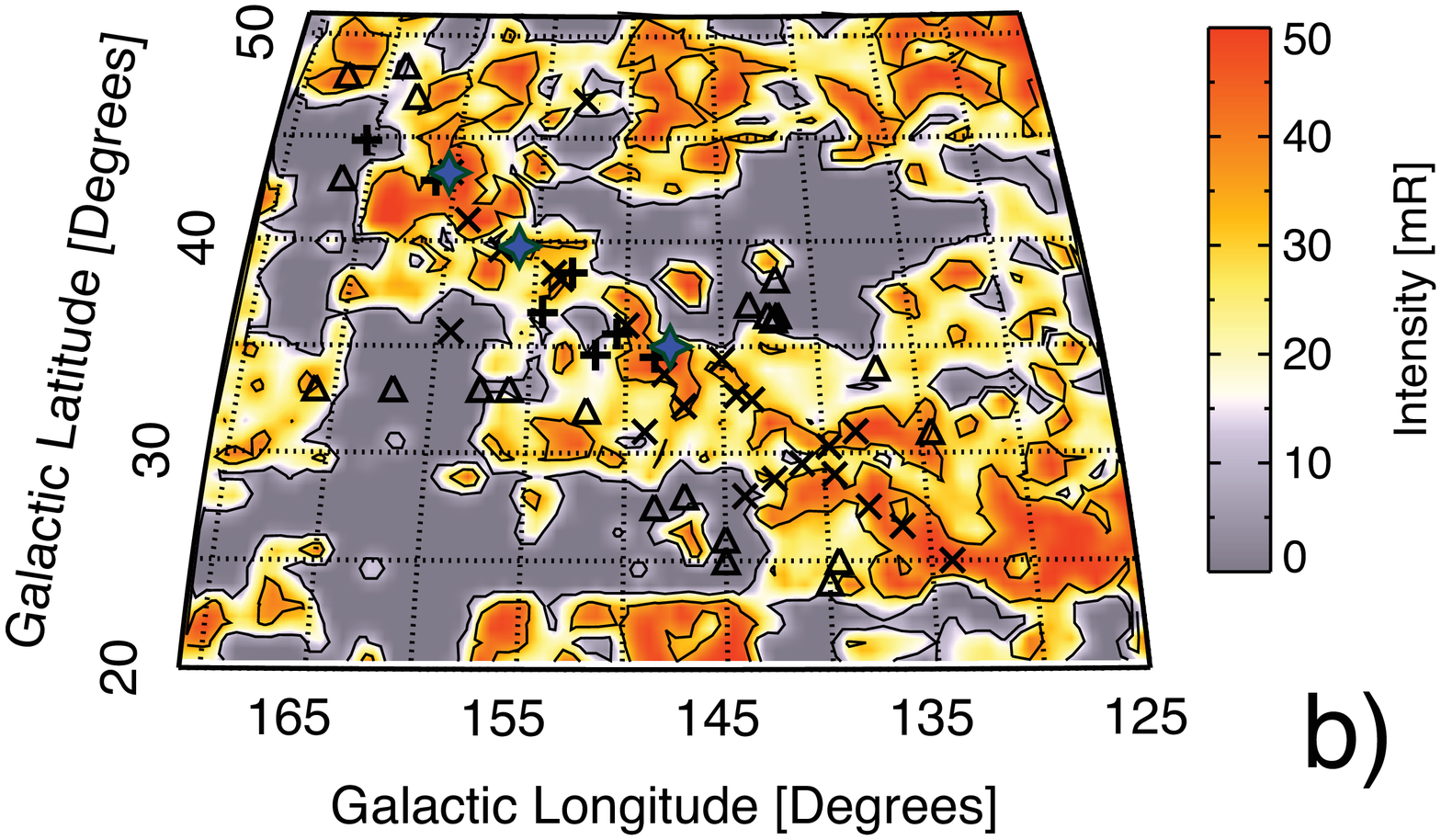} \\
\includegraphics[scale=.35,angle=90]{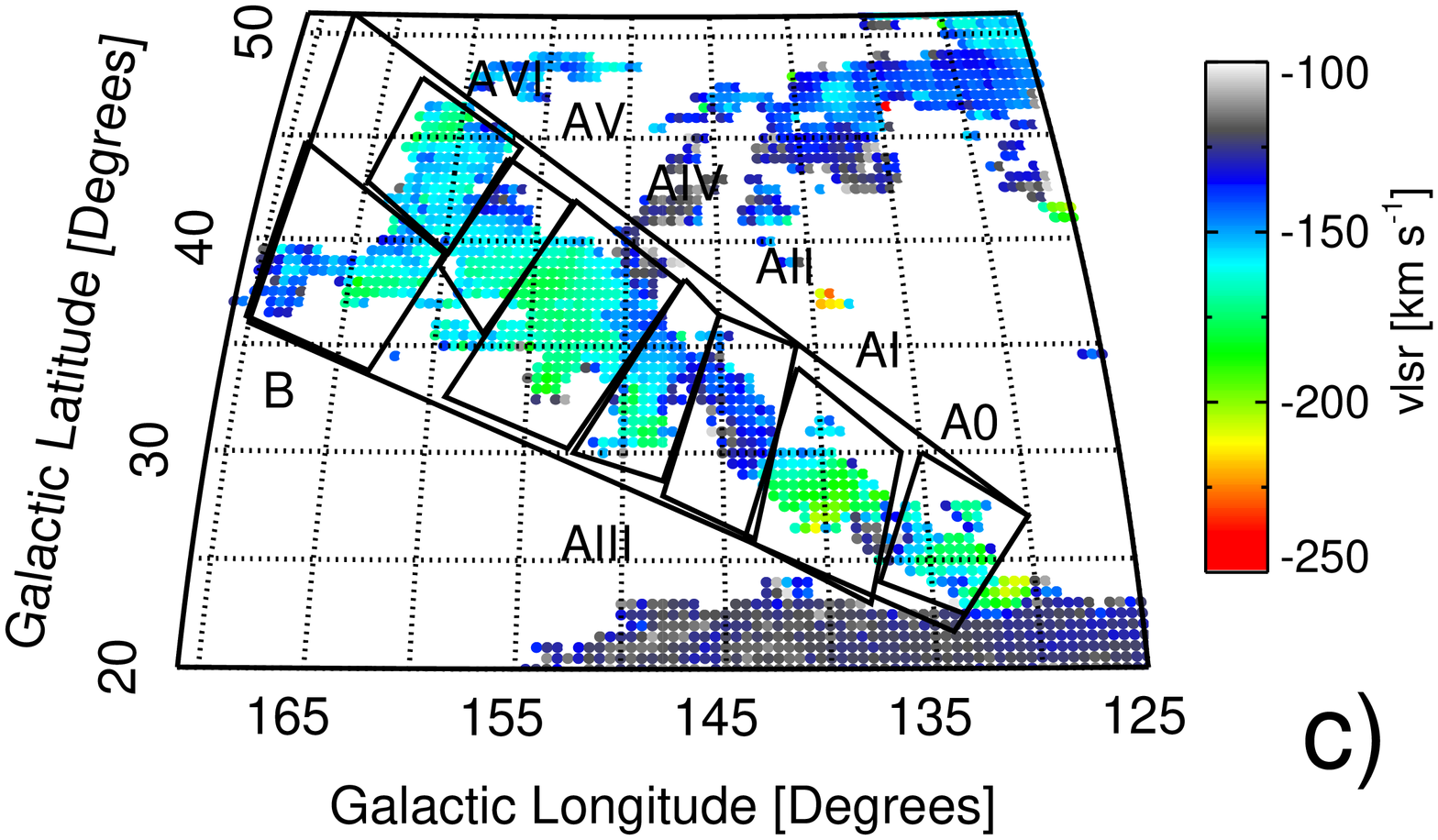}\includegraphics[scale=.35,angle=90]{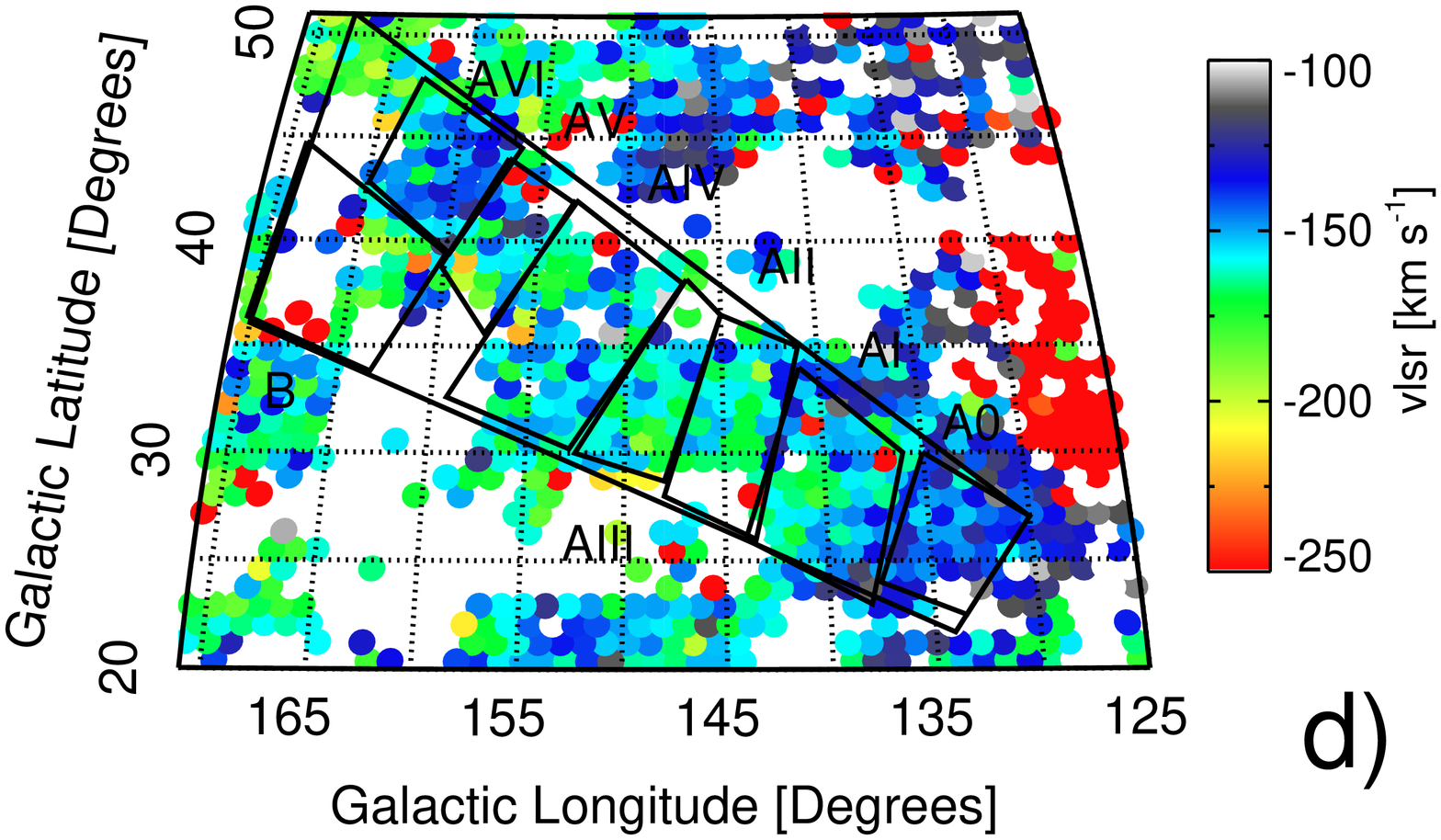} 
\figcaption{
The \hi\ \textit{(a)} and \ha\ \textit{(b)} emission maps and the corresponding \hi\ \textit{(c)} and \ha\ \textit{(d)} first moment maps. The \hi\ data comes from the LAB survey and the \ha\ from WHAM observations. The emission is integrated over the $-220$ to $-110~\kms$ \vlsr\ range. The contours in panel \emph{(a)} follow the $0.5\times10^{19}~\cm^{-2}$ and $2.5\times10^{19}~\cm^{-2}$ \hi\ emission level and the contours in panel \emph{(b)} trace the $10$ and $35~{\rm mR}$ \ha\ emission level. The cross and plus symbols represent the pointed \ha\ line and pointed \sii, \nii, \oi\ line observation locations, respectively. The triangles depict the locations of off-source observations and the diamonds show the sightlines of distance measurement studies (see Section \ref{section:distance}). The black boundaries, shown in panels \emph{(c)} and \emph{(d)}, encompass the regions used to determine the mass of Complex~A. Table \ref{table:region} lists the corner positions of these boundaries. Faint \ha\ emission more than 2.5 degrees off of the $3\times10^{18}\ \rm{cm}^2$ \hi\ contour was excluded, as shown, in determining the mass of each region because a definitive spectroscopic association could not be made between the faint \hi\ and \ha\ gas. 
\label{figure:Map}}
\end{center}
\end{figure*}

Numerous 21 cm studies have provided insight on the neutral gas component of HVCs, but few studies have investigated their warm $(10^4~{\rm K})$, ionized gas phase. This is because the emission lines from their ionized components are usually very faint, often less than a tenth of a Rayleigh,\footnote{$1~\rm{Rayleigh} = 10^6 / 4 \pi \textrm{ photons} \cm^{-2} \sr^{-1} \s^{-1}$, which is $\sim1.7\times10^{-6}\,{\rm erg} \cm^{-2}\s^{-1}\sr^{-1}$ at \ha.} and require a highly sensitive instrument to detect  (e.g., Complex K: \citealt{2001ApJ...556L..33H}, Complex L: \citealt{2005ASPC..331...25H}, Smith Cloud: \citealt{1998MNRAS.299..611B, 2003ApJ...597..948P, 2009ApJ...703.1832H}, Complexes M, A, C: \citealt{1998ApJ...504..773T}). Absorption-line studies have demonstrated the existence of diffuse ionized gas in the circumgalactic medium in both our local galactic neighbors and in higher redshift galaxies (e.g., \citealt{2003ApJS..146..165S, 2005ASPC..331...11W, 2006ApJS..165..229F, 2009ApJ...699..754S, 2009ApJS..182..378W}, \citealt{2007MNRAS.377..687L, 2011Sci...334..955L}). However, because the sampling depends on the availability of background objects, these studies cannot quantify the large-scale distribution, morphology, and total mass of the gas. 

Several studies place the heliocentric distance to Complex~A between $8$ and $10~\kpc$, or $5$ to $7~\kpc$ above the Galactic plane (\citealt{1996ApJ...473..834W,1997MNRAS.289..986R, 1999Natur.400..138V, 2003ApJS..146....1W}). This distance suggests an \hi\ mass of $\sim10^6\, M_{\odot}$ \citep{2004ASSL..312..195V}. There are currently no estimates for the mass of the warm ionized gas. \citet{1998ApJ...504..773T} identified ionized gas in Complex~A towards cores~AIII and AIV with \ha\ detections of $80\pm10$ and $90\pm10$~mR, respectively, indicating the existence of a warm ionized component. Studies have found a sub-solar composition for Complex~A (\citealt{1994A&A...282..709K, 1995A&A...302..364S, 1996ApJ...473..834W, 1999Natur.400..138V, 2001ApJS..136..537W}), which suggests that it is new material accreting onto the Galaxy. Quantifying how much ionized gas is in Complex~A is essential to understanding how this material will affect the Milky Way and the physical conditions of the gas in the Galactic halo. 

\begin{deluxetable*}{lccccc}
\tabletypesize{\scriptsize}
\tablecaption{Region Locations\label{table:region}} 
\tablewidth{200pt}
\tablehead{
\colhead{Region}                            	& \colhead{sq. deg}			& \colhead{(\textit{l,b})$_1$}	  		& \colhead{(\textit{l,b})$_2$} 				& \colhead{(\textit{l,b})$_3$} 			& \colhead{(\textit{l,b})$_4$}  }
\startdata				  
Total 						& 433.5					& 168\fdg0, 51\fdg0 					&130\fdg0, 27\fdg0 						&134\fdg1, 21\fdg7 					&170\fdg0, 36\fdg0 	\\	
A0							& 30.4					& 135\fdg0, 30\fdg0 				 	&130\fdg0, 27\fdg0 						&133\fdg5, 22\fdg5 					&137\fdg5, 24\fdg0 	\\
AI							& 50.0					& 141\fdg0, 34\fdg0 					&136\fdg0, 30\fdg0 						&138\fdg0, 23\fdg0 					&144\fdg0, 26\fdg0 	\\
AII							& 42.0					& 145\fdg0, 36\fdg5 			 		&141\fdg0, 35\fdg0 						&143\fdg5, 26\fdg0 					&148\fdg0, 28\fdg0 	\\
AIII							& 31.7					& 146\fdg8, 38\fdg2 			 		&145\fdg0, 36\fdg5 						&148\fdg0, 28\fdg7 					&152\fdg5, 30\fdg0 	\\
AIV							& 72.5					& 153\fdg0, 42\fdg0 				 	&147\fdg0, 38\fdg0 						&152\fdg8, 30\fdg2 					&159\fdg2, 32\fdg6 	\\
AV							& 29.5					& 157\fdg0, 44\fdg0 				 	&153\fdg0, 41\fdg8 						&157\fdg5, 35\fdg6 					&160\fdg2, 38\fdg8 	\\
AVI							& 39.1					& 162\fdg8, 47\fdg9 			 		&156\fdg3, 44\fdg5 						&159\fdg8, 39\fdg4 					&164\fdg9, 42\fdg8 	\\
B	 						& 64.4					& 169\fdg0, 44\fdg7 					&160\fdg0, 39\fdg4 						&163\fdg3, 33\fdg8 					&170\fdg2, 36\fdg3 	\\	
\enddata
\tablecomments{The (\textit{l,b})$_{n}$ indicates the corner positions for the regions and are shown graphically in Figures \ref{figure:Map}.}
\end{deluxetable*}

Ionizing radiation from the Milky Way and extragalactic background could ionize Complex~A, but some of the ionization could also come from interactions with the Galactic halo. If photoionization is predominantly responsible for producing the \ha\ emission, then measuring the \ha\ emission will constrain the escaping Lyman continuum flux from the Milky Way. If interactions with the halo produce \ha\ emission, then this cloud would probe the physical conditions of the surrounding halo gas. The unique elongated morphology of Complex~A enables an investigation of these processes across multiple kiloparsecs. Mapping the \ha\ emission along the entire length of this complex will aid in determining the source of the ionized gas.  

This study quantifies the extent, the amount, and the properties of the warm gas in Complex~A through faint emission lines, allowing us to understand how accretion of this $\sim10^6\, M_{\odot}$ of material will impact the metallicity and star formation rate of the Milky Way. In addition, we investigate the source of ionization and where this cloud and---similar clouds---could have formed. Section \ref{section:obs} describes mapped \ha~and pointed \ha, \sii $\lambda6716$, \nii $\lambda6584$, and [O~{\sc i}]$\lambda6300$ observations. In Section \ref{section:modeling}, we define the modeling parameters used to interpret  the observations, including the \ha\ extinction correction, the distance to the cloud, and the neutral and ionized gas distribution. Section \ref{section:ionization} investigates if photoionization modeling from the Milky Way and extragalactic background can reproduce the observed \ha\ emission and produce realistic cloud properties. We then use the observations and the results from the photoionization modeling to place constraints on the physical properties of the cloud. In Section \ref{section:temperature}, we use the \ha\ line widths to constrain the electron temperature. Then, Section \ref{section:mass} investigates the total mass of the cloud by addressing the distribution of neutral and ionized gas and projection angle of the cloud. Section \ref{section:abundance} describes how our multi-wavelength detections restrict the $\mathrm{S} / \mathrm{H}$ and $\mathrm{N} / \mathrm{H}$ abundances. In Sections \ref{section:implications} and \ref{section:summary}, we conclude with a discussion of the implications of this study and a summary of our results.

\section{Observations}\label{section:obs}

Using the Wisconsin \ha\ Mapper (WHAM) at the Kitt Peak National Observatory, we observed Complex~A in 2005 and 2006 at the Kitt Peak National Observatory. WHAM is optimized to detect faint optical emission from diffuse ionized sources. The spectrometer, described in detail by \citet{2003ApJS..149..405H}, consists of a dual-etalon Fabry-Perot that produces a $200~\kms$ wide spectrum with $12~\kms$resolution integrated over its $1\arcdeg$~beam---the emission within the 1\arcdeg~beam is averaged to produce one spectrum. For this study, we configured the spectrometer to detect line emission in the range of $-250\lesssim \vlsr \lesssim -50\ \kms$ in both mapped and pointed modes, described below, where $\vlsr$ is the velocity in the local standard of rest frame.

\subsection{Mapped Observations}\label{sec:mapping}

WHAM surveyed a large region of the Galactic halo from $(\mathit{l, b}) = (124\arcdeg, 18\arcdeg)$ to $(171\arcdeg, 53\arcdeg)$ and from $-250$ to $-50~\kms$ using the strategy outlined by \cite{2003ApJS..149..405H} for the Northern Sky Survey. As in the survey, observations are grouped into ``blocks'', 30--50 spectra mapped in a raster and taken sequentially in time. Individual exposures were 60 seconds with some locations, primarily along the length of the cloud, observed multiple times over multiple months. Throughout the year, the Earth's orbit shifts Galactic emission (fixed in the LSR velocity frame) with respect to the widespread atmospheric lines that dominate the baseline (fixed in the geocentric velocity frame). Combining observations that span many months helps extract faint emission and minimizes residuals from the removal of the atmospheric background. 

Figure \ref{figure:MapSpectraSample} shows a representative series of spectra taken over three nights toward two directions at a similar Galactic latitude ($b = -34\arcdeg$). Spectra in panel \emph{(a)} were obtained toward core~AIII, a region of higher \hi\ column density as mapped by 21~cm observations. Panel \emph{(b)} shows  spectra from a direction near, but off the neutral Complex. In the top panels, we plot unreduced data against a geocentric velocity frame where zero corresponds to the wavelength of the \ha\ recombination line. Atmospheric features align vertically from night to night in this frame of reference while Galactic features may move between observations. The wide feature near $v_{geo} = -210$ \kms\ is ``ghost'' emission from the bright \ha\ geocorona at $v_{geo} = -2.3$ \kms\ due to incomplete suppression of a neighboring order in the high-resolution etalon (see \citealt{2003ApJS..149..405H}, Figure 2). The rise in the spectra near the red edge is due to lower-velocity emission from the Galaxy in this direction. Remaining ripples in the baseline between these two figures arise from faint atmospheric lines ($\rm{I}_{\mha} \lesssim 0.2$~R) and high-velocity Galactic \ha, our target of study. Different sky conditions and zenith distances cause a slight shift in vertical displacement in the baseline continuum in these spectra.

\begin{figure*}
\begin{center}
\includegraphics[scale=.6,angle=0]{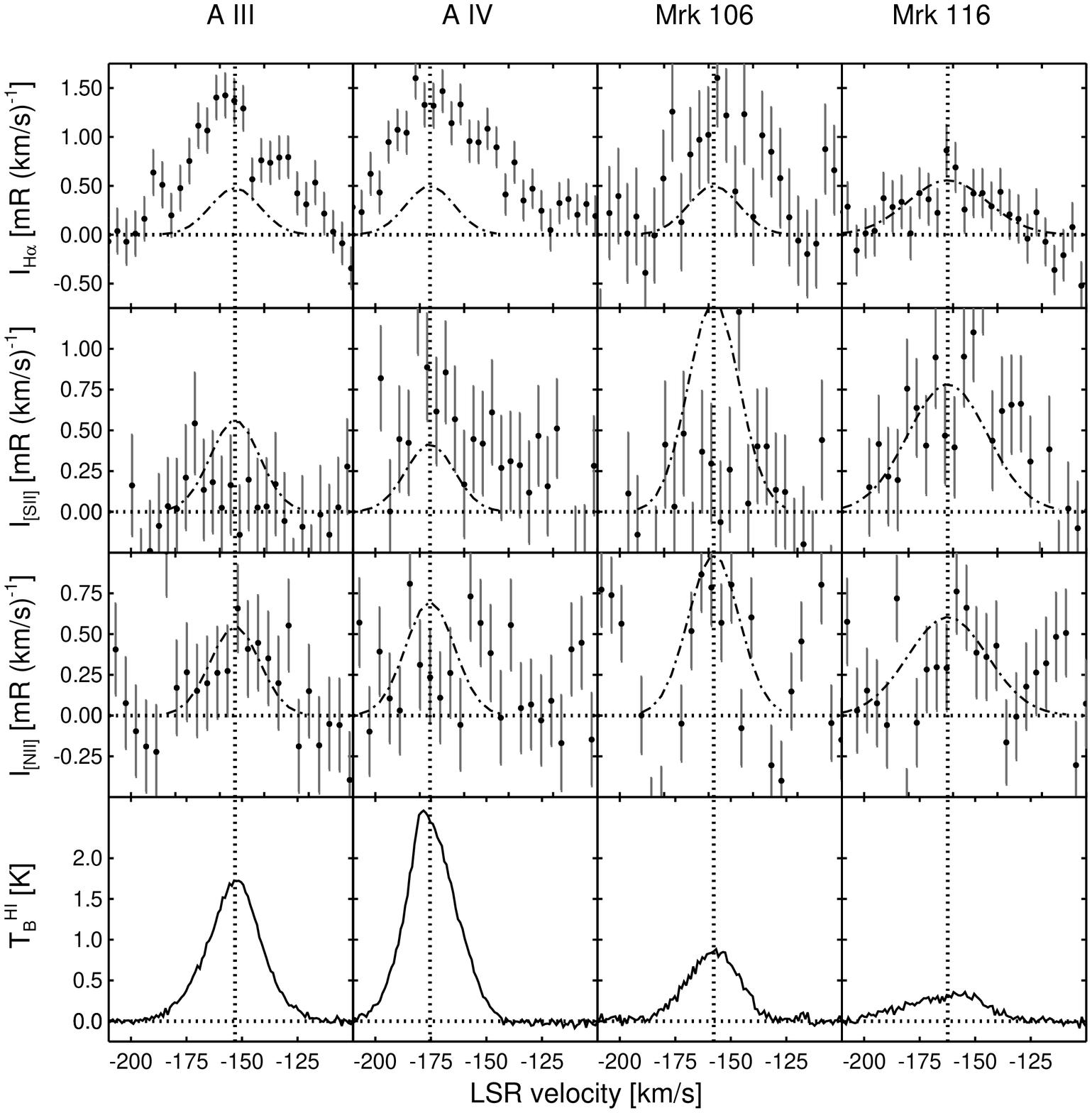}
\end{center}
\figcaption{
Multiwavelength spectra for the sightlines slightly off of core~AIII at $(l,b)=(148\fdg5, 34\fdg5)$ and core~AIV at $(l,b)=(153\fdg0, 38\fdg5)$ and towards Mrk~106 at $(l,b)=(161\fdg1, 42\fdg9)$ and Mrk~116 at $(l,b)=(165\fdg5, 44\fdg8)$. Unfavorable weather inhibited us from collecting pointed \ha\ observations of towards the Mrk~106 direction; we display the \ha\ spectrum from the mapped data set or this sightline. The core~AIII and core~AIV observations correspond to sightlines observed by \citet{1998ApJ...504..773T} in \ha. The vertical dotted line represents the center location of the H{\sc~i}~emission line and the dot-dashed line represents a Gaussian profile with the mean and width the same as the corresponding H~{\sc i} component and the area set to three times the standard deviation of the background in the same optical spectrum. Spectra are binned to $4~\kms$ to reduce systematics from faint atmospheric lines in the spectra.
\label{figure:multi_pointings}}
\end{figure*}

To extract the Galactic emission, we follow a procedure similar to \cite{2003ApJS..149..405H} for the Northern Sky Survey to subtract terrestrial emission. We construct a model of the atmospheric emission with a series of Gaussians using spectra from many regions across the sky expected to contain little to no high-velocity Galactic emission. This template, consisting of the geocentric velocity centroids, widths, and relative intensities for each of the faint atmospheric lines, is scaled and combined with a best-fit, zero-order continuum level to create the background for each individual spectrum. For this work, we have extended the atmospheric template presented in \citet{2002ApJ...565.1060H} to more negative velocities  (thin, solid line at $\vlsr > -200~\kms$ in Figures \ref{figure:MapSpectraSample}a and \ref{figure:MapSpectraSample}b). To set the scale of the template, we fit high signal-to-noise ``block averages'' and estimate the absolute intensity of the template by adjusting it to minimize the $\chi^2$ of the fit. This value is then used to scale the template as it is applied to each individual spectrum within the block.

For this dataset, an additional Gaussian for the geocoronal ``ghost'' (thin, solid component in Figures~\ref{figure:MapSpectraSample}a  and \ref{figure:MapSpectraSample}b near $v_{geo} = -210$~\kms) is needed in every observation. Finally, when warranted, we add any components for high-velocity Galactic emission (thick, solid component in Figure~\ref{figure:MapSpectraSample}a near $v_{geo} = -150~\kms$) to fully model a spectrum. The parameters for these (typically two) Gaussian components are allowed to vary to minimize the $\chi^2$ of our fit. Using these model parameters, we subtract the background continuum, atmospheric template, and ``ghost''  component from the raw data. Spectra from multiple nights toward the same direction are averaged in the LSR frame to produce the final spectra used in this work. Figure~\ref{figure:MapSpectraSample} shows a representation of this procedure for two sample directions, one toward the neutral concentration AIII and one where no 21-cm emission has been detected at the complex velocities. Panel (c) shows a distinctive component in the fully reduced spectrum from the first direction at Complex~A velocities with an intensity of 65~mR.

\begin{figure}
\includegraphics[scale=.35,angle=0]{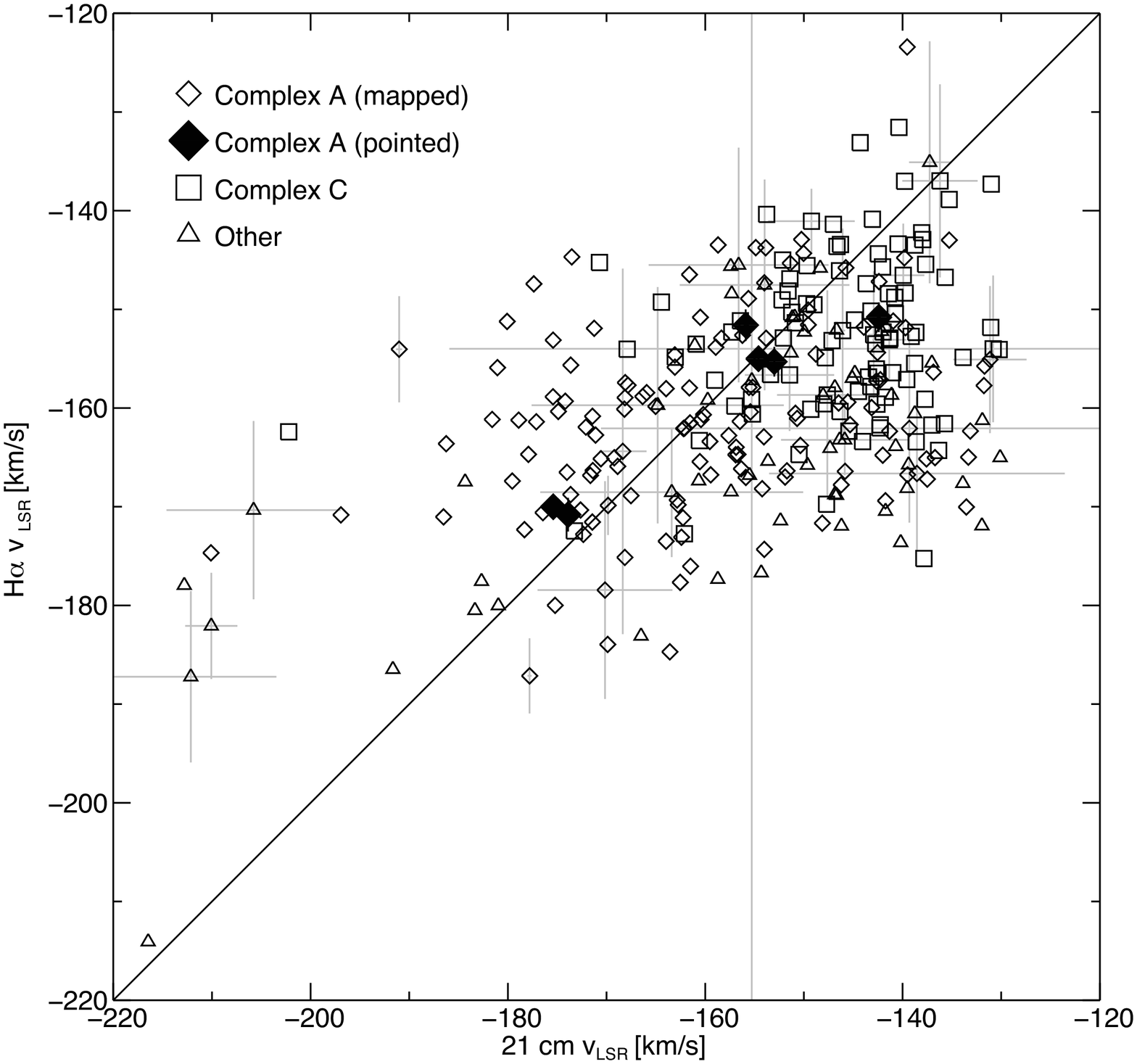} 
\figcaption{
The center of the \ha\ lines compared with the center of the \hi\ lines for the sightlines with detections. Both the filled and outlined diamonds represent Complex~A; the large black diamonds correspond to only the pointed detections with near Gaussian profiles. The squares indicate sightlines within Figure \ref{figure:Map} that include emission from Complex~C. The triangles signify sightlines that contain complementary \ha\ and \hi\ emission that could be originating from Complex~A, Complex~C, or Galactic foreground. To reduce clutter, only the 30 brightest \ha\ points include error bars. A diagonal solid line with a slope of one is provided for reference.
\label{figure:ComA_velocity}}
\end{figure}

The map shown in Figure~\ref{figure:Map} integrates this new dataset over velocities that trace Complex~A. We also display the corresponding 21~cm emission from the Leiden/Argentine/Bonn Galactic H{\sc~i}\ Survey (LAB; \citealt{2005A&A...440..775K}) over the same velocity range for comparison. This 21-cm survey will be used throughout our work to compare the \hi\ and \ha\ emission. The new \ha\ dataset contains some of the faintest extended emission regions we have mapped with WHAM to date. Our reduction method reveals \ha\ emission in the mapped region from Complex~A as well as from a portion of Complex~C, $(\mathit{l, b}) = (124\arcdeg, 45\arcdeg)$ to $(150\arcdeg, 53\arcdeg)$, and an outer arm of the Galaxy $(\mathit{l, b}) = (124\arcdeg, 18\arcdeg)$ to $(155\arcdeg, 25\arcdeg)$. Indeed, even the faintest \ha\ emission components appear at a similar velocity to that seen in the 21~cm (compared later in Figure \ref{figure:ComA_velocity}) giving us confidence in our methods. The \ha\ emission has much less contrast but a broader angular extent than the 21~cm emission. Some of this difference could be related to the different beam sizes of the \hi\ and \ha\ observations, where the WHAM beam is almost 2 times larger in diameter than the LAB survey's, and the insensitivity of the LAB survey to highly ionized regions with low $N_{\rm H{\textsc{~i}}}$. The LAB survey has a $3\sigma$ $N_{\rm H{\textsc{~i}}}$ sensitivity of $\sim2\times10^{18}~\cm^{-2}$ for typical HVCs with widths of $30~\kms$. As a result, the diffuse ionized emission further from the cores of the complexes may represent nearly fully ionized gas in the outskirts of these halo structures. The implications of this extended plasma for Complex~A are more fully discussed in Sections~\ref{section:distribution} and \ref{section:mass}.

The global velocity structure of the warm and neutral gas generally agree (see Figures \ref{figure:Map}c--\ref{figure:Map}d), but the \ha\ gas has a lower mean velocity at core A0, higher at core AII, and lower at core AVI. The Galactic foreground \ha\ emission towards core A0 causes confusion, which has lowered the mean velocity towards this core. Both of the \hi\ and \ha\ first-moment maps of Complex~A, or intensity-weighted mean velocity along the line of sight, indicate that neither the warm or neutral velocity components behave monotonically from core A0 to core AVI. %Complex~C, in the upper right corner of Figures \ref{figure:Map}, also has a mean velocity similar to Complex~A in \ha, which could mean that their ionized gas components might be associated with each other.

\subsection{Pointed Observations}\label{section:pointed}

In addition to mapping Complex~A in \ha, we obtained multiple, deep, 120-second ``pointed'' observations in \ha, \sii, \nii, and \oi. In Table \ref{table:intensity_time} at the bottom of this article, we list the non-extinction-corrected intensities, the integrated exposure times, and the dates observed for the pointed observations. For observations taken towards Mrk~106, PG~0822+645, PG~0832+675, PG~0836+619, and IRAS~08339+6517, weather prevented the collection of pointed \ha\ observations. For these sightlines, Table \ref{table:intensity_time} lists intensities from the mapped observations. The atmospheric contamination in the pointed observations was removed by subtracting the spectrum of a nearby off-source direction. We remove any residual continuum in the atmospheric subtracted spectrum by fitting a constant background level.  

Although the pointed and mapped observations have different total exposure times and underwent atmospheric subtraction through a different procedure, many of the \ha\ intensities agree to within the combined uncertainties in overlapping pointed and mapping observations; however, the pointed observations are prone to increase uncertainty due to systematic effects associated with minor atmospheric differences between the ons and offs. These differences can cause notable discrepancies between the intensities derived from these two methods. Two main effects increase the noise of the pointed observations: (1) Some of the offs contain trace amounts of ionized gas emission. The patchy nature of the ionized gas emission increases the probability that the offs partially overlap with regions containing ionized gas, especially with 1-degree angular resolution of these observations. Further, the ionized component of Complex~A may extend many degrees off of the neutral component as typical with other HVCs (e.g., \citealt{2012MNRAS.424.2896L}). (2) The atmospheric contribution varies slightly from position to position on the sky, which can significantly decrease the signal-to-noise ratio of the resultant spectra when observing extremely faint objects like Complex~A. For atmospherically subtracted pointed spectra with ill-defined spectral components, we only report upper limits in Table \ref{table:intensity_time}. The mapped \ha\ observations presented in Figure \ref{figure:Map} do not suffer from the same effects as the pointed values as they are reduced using an atmospheric template and are therefore more reliable. 

\begin{figure}
\begin{center}
\includegraphics[scale=.4,angle=90]{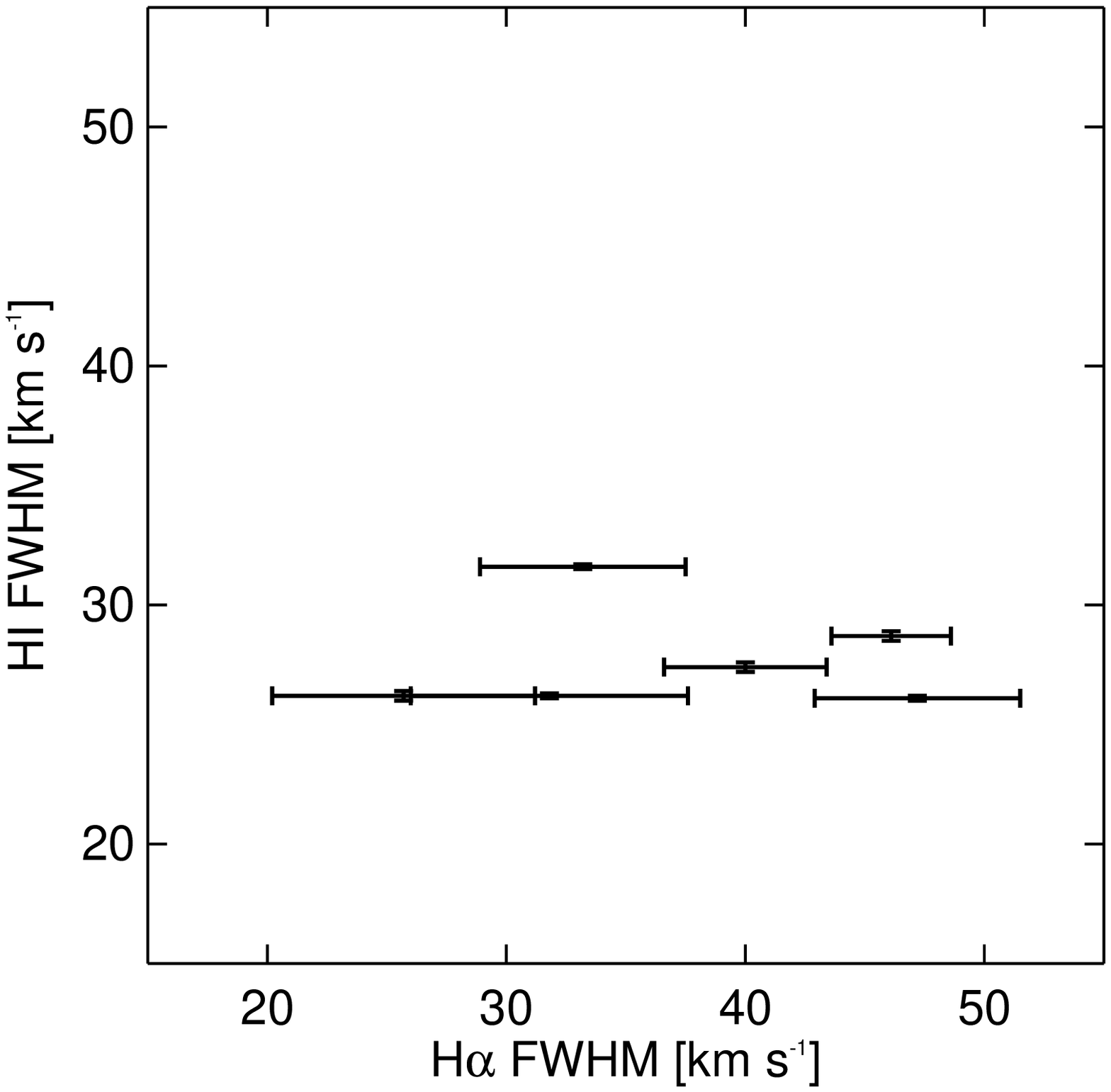}
\end{center}
\figcaption{
The full-width at half-maximum of the H{\sc~i}\ and \ha\ lines for the pointed sightlines with nearly Gaussian profiles. 
\label{figure:ComA_fwhm}}
\end{figure}

These optical emission-line spectra include emission from the Galactic warm ionized medium, especially in \ha, \sii, and \nii\ spectra at velocities $>-120~\kms$. The \ha\ and the \oi\ suffer from an additional spectral contamination at velocities $<-220~\kms$~in the form of an incomplete suppression of a neighboring high-resolution etalon order of the geocoronal line. For these reasons, we avoided the \vlsr\ $> -120$ and \vlsr\ $<-220~\kms$~spectral regions when determining the background level. In addition, the \oi\ spectra include faint, time-varying, atmospheric lines. As such, Table \ref{table:intensity_time} only lists the upper intensity limit of the \oi\ observations. The detection limit is the calculated area of a Gaussian with a width equal to that of the \hi\ line along the same sightline and a height equal to 3 times the standard deviation of the scatter in the background. To account for the systematic uncertainties from the faint atmospheric lines, in addition to reporting the statistical uncertainties, Table \ref{table:intensity_time} also reports systematic uncertainty---the uncertainty of the measurement. To characterize both the statistical and systematic uncertainties, we fit each spectrum multiple times. First, we calculated the value which minimizes $\chi^2$ with the mean, width, and the area of the Gaussian unconstrained; this yielded the reported intensity and the statistical error. Second, we varied the base line of the fit to attain $\Delta \chi^2 = 1$. Third, we fit the line by constraining various fitting parameters, including the mean velocity and width---measured from the average \hi\ spectra along the same sightline---held constant; the \hi\ spectra were smoothed to a $1\arcdeg$ angular resolution to match the \ha\ resolution. Parameters describing the \ha\ fit with \hi\ mean velocities and width along the same sightline were incorporated because of the asymmetrical profile of the \ha\ emission. The systematic uncertainties of the \ha\ intensity reported in Table \ref{table:intensity_time} are the root-mean-square of the intensities derived from these fits.

\begin{figure}
\includegraphics[scale=.3,angle=0]{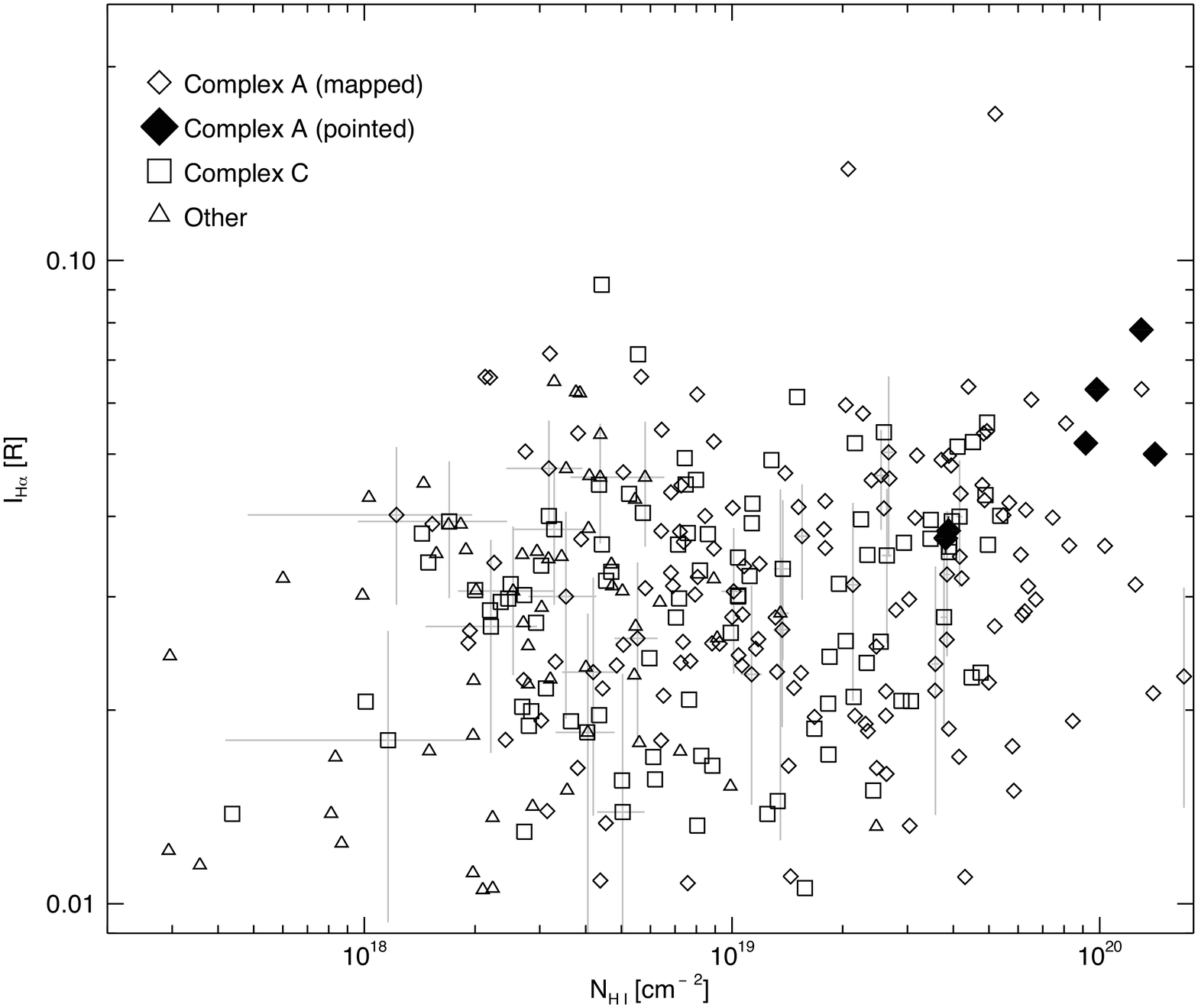} 
\figcaption{
The \ha\ intensity compared with the \hi\ column density for sightlines with detections in both lines. Both the filled and outlined diamonds represent Complex~A; the large black diamonds correspond to only the pointed detections with near Gaussian profiles. The squares indicate sightlines within Figure \ref{figure:Map} that include emission from Complex~C. The triangles signify sightlines that contain complementary \ha\ and \hi\ emission that could have originating from Complex~A, Complex~C, or Galactic foreground. To avoid clutter, one in ten points---chosen randomly---include an error bar.
\label{figure:ComA_intensity}}
\end{figure}

The \nii\ and \sii\ emission lines from the ionized gas in Complex A are faint. In many directions, the emission is below our $3 \sigma$ detection limit. The sightlines with signal above the $3\sigma$ detection limit  are listed in Table \ref{table:intensity_time}. To visually represent this upper limit, Figure \ref{figure:multi_pointings} includes $3 \sigma$ Gaussian profiles. Figure \ref{figure:multi_pointings} displays four sightlines with \ha\ detections, but only the core~AIII and Mrk~106 sightline have detectable \nii\ emission and only core~AIV and Mrk~116 sightlines have detectable \sii\ emission. The construction of this ``ideal'' three sigma profile incorporated two assumptions for the spectra of the neutral and ionized gas along the same sightline: (1) that they have the same peak velocity and (2) that they have the same width. The height of the ideal profile is defined as 3 times the standard deviation of the background. Although this is an idealized profile, the center velocity of \hi\ and the \ha\ lines generally agree in HVCs (See Figure \ref{figure:ComA_velocity}; \citealt{1998MNRAS.299..611B, 1998ApJ...504..773T, 2003ApJ...597..948P, 2009ApJ...703.1832H}). Since the \ha\ width cannot be measured directly towards sightlines with an asymmetrical distribution, the second assumption provides an estimate of the minimum width expected. Figure \ref{figure:ComA_fwhm} shows that the \ha\ lines can be substantially wider than the \hi\ lines. Wider \ha\ lines could indicate that the ionized gas is warmer, that non-thermal line-broadening effects influence the ionized gas more than the neutral gas, or that the ionized gas is present in multiple, distinct components along the line of sight. Broad \ha\ lines are certainly consistent with a scenario where photoionization provides an additional heating source for the gas. However, bulk dynamic effects, such as enhanced internal turbulence or expansion of an ionized region, could also account for the difference in widths. Finally, the \ha\ emission may be exposing regions of the complex that are highly ionized and have become kinematically distinct from its neutral core ($\mathrm{N_{H\,I}} \gtrsim 10^{17}~\cm^{-2}$). In any case, the combination of the lack of correlation between the \ha\ and \hi\ widths and intensities, as shown in Figure \ref{figure:ComA_intensity}, suggests that the \ha-emitting gas is likely not mixed with the \hi\ that shapes the 21-cm profiles.

Many of the pointed \ha\ observations targeted dense \hi\ cores. We obtained additional multiline spectra slightly offset from neutral cores AIII and AIV to follow-up on the \ha\ detections in \citet{1998ApJ...504..773T}. The \ha\ intensities reported here differ from those listed in \citet{1998ApJ...504..773T}, $80\pm10~{\rm mR}$ for core~AIII and $90\pm10~{\rm mR}$ for core~AIV, as a consequence of using constant baselines instead of sloped ones. Interestingly, these sightlines show brighter \ha\ emission than the sightlines centered on the \hi\ cores. Figure \ref{figure:multi_pointings} displays the spectra for these two sightlines. Many of the combined spectra taken over the two month observing cycle towards faint directions became flattened, broadened, or an artificial multiple peaked distribution due to shifted atmospheric lines in the local standard of rest frame.
%The shift of the local standard of rest frame with respect to the geocentric frame over the two month observing period caused many of the combined spectra towards faint directions to flattened and broadened due to the shifted atmospheric lines. 

\begin{comment}
To investigate how the \ha\ emission varies along the length of Complex~A, we obtained pointed \ha\ observations along the length of the cloud at different Galactic latitudes (A$_{31~\rm{to}~41}$)\footnote{The subscript indicates the latitude of the observation.}. These sightlines avoided the \hi\ cores. No obvious pattern emerged in the emission-line strengths. These sightlines demonstrate that the \ha\ emission is patchy; the large scale structure of \ha~traces the \hi, but these emission lines seem uncorrelated at small scales as demonstrated in Figure \ref{figure:ComA_intensity}, which compares the \hi\ intensity to the \ha.
\end{comment}

To complement past and future absorption-line work, many of the pointed sightlines were targeted towards sightlines with known background quasars and UV-bright stars. UV absorption spectra towards background quasars constrain the metallicity of the absorbing HVCs; absorption towards foreground and background stars with known distances, such as RR Lyrae and blue horizontal branch stars, have been used to bracket the distance to the HVC (see \citealt{2001ApJS..136..463W} and \citealt{2008ApJ...672..298W} for compilations of these studies). The noisy Mrk~116 spectra in Figure \ref{figure:multi_pointings} suggests the presence of ionized emission, but the \ha\ emission is below the $3 \sigma$ detection limit. A majority of the \oi\ spectra also demonstrate this two-component behavior; this is likely due to contamination from small changes in the atmosphere during the observations.

\section{Modeling Parameters}\label{section:modeling}

These new observations of Complex~A place important constraints on some of the physical properties of the HVC and its immediate environment within the halo. In Sections \ref{section:ionization} through \ref{section:abundance}, we explore the source of ionization, electron temperature, mass, and metal abundance of Complex~A. Here we discuss several other input parameters that complement the observations which we use to model the HVC.
                       
\subsection{\ha\ Extinction Correction}\label{section:extinction}

The position of Complex~A above the Galactic Plane results in minimal interstellar dust extinction. We expect that most of the extinction comes from either local interstellar dust or dust in the foreground Perseus Arm. We use the reddening given in \citet{1994ApJ...427..274D} for a warm diffuse medium with a low-density: 
\begin{equation}
E(B-V)=\frac{\langle N_{\rm H{~\textsc{i}}}\rangle}{4.93\times10^{21}~\cm^{-2}}\ {\rm mag},
\end{equation}
where the average \hi~column density $\langle N_{\rm H{~\textsc{i}}}\rangle$ includes only the foreground \hi\ emission. If the extinction follows the $\langle A(\mha)/A(V)\rangle=0.909-0.282/R_v$ 
optical curve presented in \citet{1989ApJ...345..245C} for a diffuse ISM, where R$_v\equiv A(V)/E(B-V)=3.1$, then the expression for the total extinction becomes
\begin{equation}
A(\mha)=\frac{\langle N_{\rm H{~\textsc{i}}}\rangle}{1.95\times10^{21}~\cm^{-2}}~\mathrm{ mag}.
\end{equation} 

The average \hi\ column density of the foreground towards the dense \hi\ cores is $3.7\times10^{20}~\cm^{-2}$, yielding an 8\% average \ha\ extinction correction. All subsequent sections correct for extinction using the LAB survey \hi\ column densities, smoothed to 1\arcdeg\ resolution to match the WHAM observations, except when comparing \ha, \nii, and \sii\ intensity ratios as the difference in extinction is negligible over such a small wavelength range.

\subsection{Distance to the Cloud}\label{section:distance}

Three studies constrain the distance to Complex~A. \citet{1996ApJ...473..834W} found no evidence of absorption from Complex~A in the spectrum of the foreground star PG~0858+593 at (\textit{l,b})=(156\fdg9, 39\fdg7)---a star with a known distance of $4.0\pm1.0~\kpc$. \citet{1997MNRAS.289..986R} and \citet{2003ApJS..146....1W} constrain the distance of Complex~A to be further than $8.1~\kpc$ from the Sun and $4.7~\kpc$ above the Galactic plane due to a lack of absorption Ca{\sc~ii}, C{\sc~ii}, Si{\sc~ii}, and O{\sc~i} absorption features associated with Complex~A in the spectra of star PG~0832+675 at (\textit{l,b})=(147\fdg7, 35\fdg0). A recent absorption-line study by \citet{2012MNRAS.424.2896L} using higher resolution and higher signal-to-noise ratio observations than previous studies indicates that PG~0832+675 may actually lie within the front edge of Complex~A, suggesting that core AIII is at a distance of $8.1~\kpc$. \citet{1999Natur.400..138V} found a maximum distance to Complex~A of $9.9\pm1.0~\kpc$ towards star the AD~UMa at (\textit{l,b})=(160\fdg0, 43\fdg3), indicated with a blue diamond in Figure \ref{figure:Map}. These studies place Complex~A between $8.1$ and $(9.9\pm1)~\kpc$ away; however, the small angular separation between these measurements---compared to the total length of Complex~A---does not anchor the projection of the cloud. 

With the elongated morphology of Complex~A, each \hi\ core could lie at a different distance. For this reason, we calculate the properties of Complex~A and the surrounding medium assuming four different projections:  (1) a constant distance of 9 \kpc, (2) inclined through (147\fdg7, 35\fdg0) and (160\fdg0, 43\fdg3) at $8.1$ and $9.9~\rm{kpc}$, respectively, (3) a constant height of $z=5.7~\kpc$ that is defined by placing the midpoint of cores AIII and AVI at a distance of 9.0~\kpc, and (4) at distances constrained by matching Lyman continuum flux  from the Milky Way and the extragalactic background with flux needed to produce the observed \ha\ emission through photoionization. If PG~0832+675 lies within Complex A, as suggested by \citet{Lehner2012}, then projection (2) corresponds to the maximum linear inclination allowed such that core A0 is projected towards the Sun.  We determine the distances for projection (4) without considering the measured distance limits for Complex~A to determine if the Milky Way and the extragalactic background with flux alone can reproduce distances that match observation or if additional ionization sources are required to reproduce the \ha\ emission.  

\subsection{Distribution of the Neutral and Ionized Gas\label{section:distribution}}

The unknown morphology along the length of the cloud and the unknown distribution of the ionized gas make accurately calculating the mass of Complex~A difficult. We estimate the mass using a procedure similar to that employed by \citet{2009ApJ...703.1832H}. For simplicity, these calculations assume that the neutral gas has a line-of-sight depth $(L)$ similar to its width as listed in column 3 of Table \ref{table:mass} at the end of this article. The depth of the ionized gas depends on its distribution, which could be well-mixed with the neutral gas or surrounding the neutral gas as an ionized skin. This distribution depends on the physical processes influencing the cloud. External ionizing sources, such as the Lyman continuum from the Milky Way and the intergalactic medium, could create a photoionized skin around the cloud. Interactions between the cloud and Galactic halo could also ionize the outer layer of the cloud through collisional ionization or through photoionization from radiation emanating from shock heated gas \citep{2007ApJ...670L.109B}.

If the warm gas is distributed as an ionized skin, that skin could be either in pressure equilibrium with a small sound crossing time compared to the recombination time or in pressure imbalance with a sound crossing much larger than the recombination time. If the ionized gas is in pressure equilibrium with the neutral gas and if the gas phases are at the same temperature, then the electron density $(n_e)$ of an ionized skin equals half the neutral hydrogen density ($n_0$; \citealt{2009ApJ...703.1832H}). For the case of pressure imbalance, we choose to explore the properties of the cloud at a constant density with $n_e = n_0$. Regardless of whether the ionized skin is in pressure equilibrium with the neutral component, the emission rate of \ha\ is proportional to the recombination rate of hydrogen $\left(\alpha_{{\rm H}_{\alpha}}\right)$ and the average number of \ha\ photons produced per recombination $\left(\epsilon_{{\rm H}_{\alpha}}\right)$: 
\begin{equation}\label{eq:I_H-alpha}
\iha=\left(4\pi\right)^{-1} \int \alpha_{B}(T)\ \epsilon_{{\rm H}_{\alpha}}(T)\  n_e\ n_p\ dl_{H^+},
\end{equation}
where $n_p\approx n_e$,  $\alpha_B=2.584\times10^{-13}\ \left(T/10^4~{\rm K}\right)^{-0.806} {\cm}^3\ {\rm s}^{-1}$ \citep{1988ApJS...66..125M}, and $\epsilon_{{\rm H}_{\alpha}}(T)\approx0.46\times\left(T/10^4~{\rm K}\right)^{-0.118}$. If the electron density and temperature are constant over the emitting region, then the path length of the ionized gas is given by
\begin{equation}\label{eq:L_ion}
 L_{H^+}=2.75\ \left(\frac{T}{10^4~{\rm K}}\right)^{0.924} \left(\frac{\iha}{R}\right) \left(\frac{n_e}{\cm^{-3}}\right)^{-2}\pc.
 \end{equation} 
 
 \begin{figure}
\begin{center}
\includegraphics[scale=0.55,angle=0]{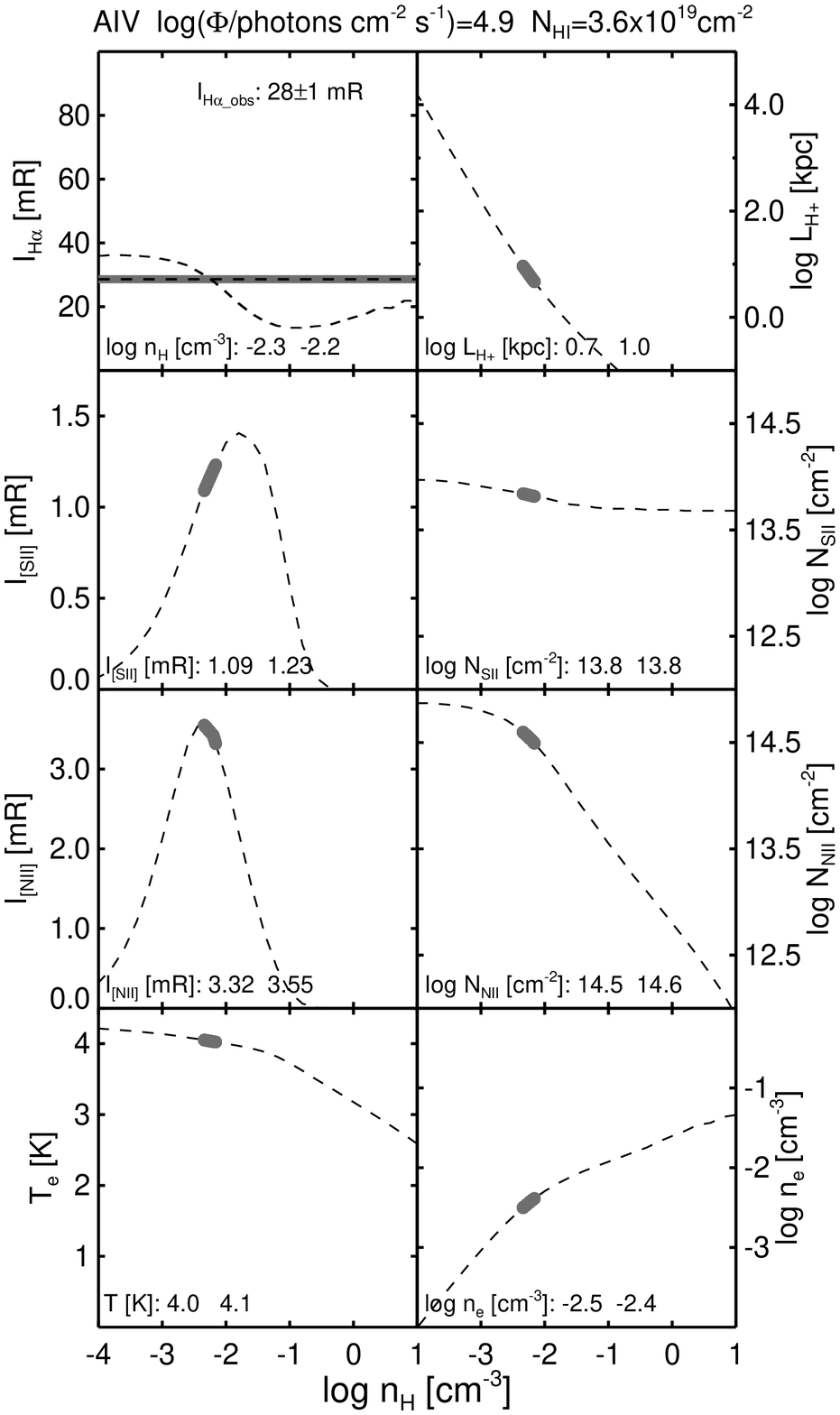}  
\figcaption{
The predicted properties of core~AIV, modeled with Cloudy, for a cloud that is purely photoionized. The incident Lyman continuum flux was determined from the \ha\ intensity with Equation (\ref{eq:FHalpha}) for clouds at $\log\left(T/{\rm K}\right) = 4.1$. The Cloudy solutions, shown as a dashed line, are found using the \hi\ column density observed in the direction of core~AIV. At low densities, asymptotic temperatures and larger path lengths compensate to give a constant \ha\ intensity. At higher densities, the temperature drops and the particles slow down; this makes recombination more efficient and causes the intensity to increase for a given emission measure. The gray horizontal envelope in the top left panel represents the average observed, extinction corrected, \ha\ intensity for core~AIV. The thick solid dark gray lines in the remaining panels trace the cloud properties for the total hydrogen densities where the predicted Cloudy \ha\ intensities match the observed \ha\ intensity. 
\label{figure:FHalpha_AIV}}
\end{center}
\end{figure}
 
 In deriving densities and temperatures, we assumed that the gas within each WHAM beam consists of three components: neutral gas with a density $n_0$ with some volume filling fraction, ionized gas with a density $n_e$ with some volume filling fraction, and an empty void filling the rest of the sightline. This approach has been long-used to describe ionized gas (e.g., \citealt{1991ApJ...372L..17R}) and is a reasonable approximation in the absence of detailed information about the distribution of the gas along each sightline. However, in reality, fluid interactions in the cloud likely produce a range of densities in both the neutral and ionized gas. If the gas is turbulent and the density distribution is lognormal, this can lead to considerably lower true mean densities and higher filling fractions than suggested by the analysis used here \citep{2008ApJ...686..363H}.

The mass calculations of Complex~A include an exploration of three possible distributions of ionized gas: an ionized skin of equal pressure to the neutral component, an ionized skin of equal density to the neutral component, and a fully mixed cloud without an ionized skin. We list the mass calculated with each of the assumptions that yield realistic cloud properties in Table \ref{table:mass} for each of the high \hi\ column density core regions.

\section{Source of the Ionization}\label{section:ionization}

A substantial amount of the gas in Complex~A is ionized, but what causes this ionization? Complex~A lies within the Galactic halo at 2.6 to 6.8~\kpc\ above the Galactic plane, depending on the orientation of the cloud. Photoionization from hot stars in the disk of the Milky Way and the extragalactic background could produce a significant amount of ionized gas, but  hydrodynamical interactions between the cloud and the Galactic halo could also collisionally ionize the cloud. To explore whether photoionization could be responsible for the observed \ha\ emission, we compare the properties of a cloud bathed in enough ionizing photons to produce the observed \ha\ emission, constrained by the observed geometrical properties of the cloud and by the measured \hi\ column density. We then compare these results with a cloud illuminated by the Milky Way and the extragalactic background to test if the these sources can produce the observed \ha\ emission. 

We model the radiative processes with Cloudy (version C08.00; \citealt{1998PASP..110..761F}) by approximating Complex~A as a plane-parallel slab with the radiation incident from one side of the cloud. Although the extragalactic background radiation is isotropic, the typical \hi\ column density of Complex~A exceeds $10^{18}~\cm^{-2}$ and therefore is optically thick to Lyman continuum radiation. For this modeling, we use an abundance of 0.1 solar for Complex~A (e.g., \citealt{2004ASSL..312..195V}; Wakker \& Hernandez 2012, in prep) and the results from \citet{2009ARA&A..47..481A} for the relative abundances. With this framework, we use Cloudy to model the physical conditions in the cloud for a series of volume density values. The thickness of the cloud is constrained by requiring the total column density of neutral hydrogen to match the observed values. Cloudy then finds the column densities of all the elements for a situation where there is an equilibrium between ionization and recombination. It also yields the temperature of the gas corresponding to the heating provided by the incoming ionizing photons. For each assumed volume density, we then use the resulting thickness and electron density to calculate the \ha\ emission measure. We estimate the incident ionizing flux by combining the extragalactic radiation field of \citet{2001cghr.confE..64H}  with the model of \citet{1999ApJ...510L..33B, 2001ApJ...550L.231B}, as updated by \citet{2005ApJ...630..332F}; this model predicts the intensity of interstellar radiation as function of wavelength and location in the Milky Way halo. Since the \hi\ column density is fixed for each HVC core, we have only a two parameter fit: the assumed projection geometry of Complex A and the volume density. We show below that matching the modeled \ha\ emission to the observed value also fixes the volume density.

If photoionization produces the observed \ha\ emission, a line produced through recombination, then the rate of total hydrogen recombination must be proportional to the incident flux of the Lyman continuum: 
\begin{align} 
\phi_{\rm{LC}} &= \int\alpha_B(T)\ n_e n_p dl_{H^+}\\ 
                          &\simeq4\pi\frac{\iha(T)}{\epsilon_{{\rm H}_{\alpha}}(T)},\label{eq:phi}
\end{align}
Combining  Equation (\ref{eq:phi}) with ${\epsilon_{{\rm H}_{\alpha}}}$ (see section \ref{section:distribution}) for a gas optically thick to Lyman continuum radiation, as the \hi\ column density is greater than $10^{18}\ {\cm}^{-2}$, the flux reduces to 
\begin{equation}\label{eq:FHalpha}
\phi_{\rm{LC}} = 2.1\times10^5\left(\frac{\iha}{0.1{\rm R}}\right)\left(\frac{T}{10^4 K}\right)^{0.118} {\rm photons}\ {\rm cm}^{-2}\ {\rm s}^{-1}.
\end{equation}
For warm gas at $10^4 {\rm K}$, heated only through photoionization, the observed \ha\ emission from Complex A corresponds to an incident ionizing flux of  $\sim10^{5}\ {\rm photons} \cm^{-2} \s^{-1}$. To explore the properties of a cloud illuminated by this ionizing flux, we model the photoionization of multiple representative clouds with Cloudy. We investigate the properties of clouds with different total hydrogen densities, $-4.0 < \log{\left(n_H/\cm^{-3}\right)} < 1.0$, guided by the observed \hi\ column densities. Requiring the observed and predicted \ha\ emission to match, as shown in Figure \ref{figure:FHalpha_AIV} for core~AIV, we compared the properties of the resultant clouds. The predicted \ha\ emission is derived using Equation (\ref{eq:L_ion}), such that $L_{H^+}=N_e/\left(n_H\times{\rm volume\ filling\ fraction\ of\ ionized\ H}\right)$, where $N_e$ is the electron column density and $n_H$ is the total hydrogen number density. Table \ref{table:flux_halpha} lists the predicted properties of Complex~A using this model. 

\begin{deluxetable*}{lcccccccccccc}
\tabletypesize{\scriptsize}
\tablecaption{Properties of Photoionized Regions with $\phi_{LC}(\textrm{\ha})$ estimated from \ha\ Observations\label{table:flux_halpha}} 
\tablewidth{0pt}
\tablehead{
\colhead{}			& \multicolumn{2}{c}{Observed}		&\colhead{}	& \colhead{}	& \colhead{}	& \multicolumn{7}{c}{Predicted}  \\
\cline{2-3} \cline{7-13}\\
\colhead{Region}	& \colhead{$\langle\log\ $\nhi$\rangle$\tablenotemark{a}}	& \colhead{$\langle\mha\rangle$\tablenotemark{a,b}}	&\colhead{}	& \colhead{$\log\phi_{LC}(\textrm{\ha})$\tablenotemark{c}}	& \colhead{} & \colhead{$\log n_H$\tablenotemark{d}} 				& \colhead{$\log n_e$\tablenotemark{d}} 	& \colhead{$\log N_{H^+}$\tablenotemark{d}} 				& \colhead{$\log T_e$\tablenotemark{d}}		& \colhead{$\log L_{H^+}$\tablenotemark{d}} & \colhead{$\log L_{{\rm{H}\textsc{ i}}}$\tablenotemark{d}} & \colhead{$d$\tablenotemark{f} }\\
\colhead{}			& \colhead{[$\cm^{-2}$]}	& \colhead{[$mR$]}				&\colhead{}	& \colhead{[$\# \cm^{-2} \s^{-1}$]}	& \colhead{} &\colhead{[$\cm^{-3}$]} 	 & \colhead{[$\cm^{-3}$]} 	&  \colhead{[$\cm^{-2}$]} & \colhead{[$K$]}	& \colhead{[kpc]} & \colhead{[\kpc]} & \colhead{[\kpc]}  }
\startdata				  
A0			& 19.4		& 52$\pm$2 		&& 5.1	&& -2.8 to -2.3 		& -2.8 to -2.4	& 20.6 to 20.0	& 4.1 to 4.1	& 2.0 to 0.9	& 2.0 to 0.9	& 6.3 to 6.5 \\
AI			& 19.6		& 42$\pm$1		&& 5.0	&& -2.6 to -2.3		& -2.6 to -2.5 	& 20.2 to 20.0	& 4.1 to 4.1	& 1.5 to 1.0	& 1.4 to 1.0	& 6.8 to 8.1 \\
AII			& 19.3		& 27$\pm$2		&& 4.9	&& -2.1 to -1.9		& -2.3 to -2.1	& 19.6 to 19.4	& 4.0 to 4.0	& 0.4 to 0.1	& 0.4 to 0.0 	& 8.5 to 10.3\tablenotemark{g} \\
AIII			& 19.6		& 36$\pm$2		&& 5.0	&& -2.3 to -2.2		& -2.6 to -2.4	& 20.1 to 20.0	& 4.1 to 4.1	& 1.3 to 0.8	& 1.2 to 0.8 	& 8.0 to 8.6 \\
AIV			& 19.6		& 29$\pm$1		&& 4.9	&& -2.6 to -2.3		& -2.5 to -2.4	& 19.9 to 19.7	& 4.1 to 4.0	& 1.0 to 0.7	& 0.9 to 0.6	& 8.6 to 11.3\\
AV			& 19.4		& 36$\pm$3		&& 5.0	&& -2.4 to -1.9		& -2.5 to -2.1	& 20.0 to 19.6	& 4.1 to 4.0	& 1.1 to 0.2	& 1.0 to 0.2	& 8.3 to 11.1 \tablenotemark{g}\\
AVI			& 19.4		& 45$\pm$2		&& 5.1	&& -2.4 to -2.1		& -2.5 to -2.2	& 20.1 to 19.8	& 4.1 to 4.0	& 1.2 to 0.5	& 1.2 to 0.5	&6.7 to 9.3 \tablenotemark{g}\\
B	 		& 19.2		& 30$\pm$3		&& 4.9	&& -2.0 to -1.6		& -2.1 to -1.9	& 19.5 to 19.2	& 4.0 to 4.0	& 0.3 to -0.3	& 0.2 to -0.4	&8.5 to 10.1\\	
\enddata
\tablenotetext{a}{Average value for the region.}
\tablenotetext{b}{Extinction corrected.}
\tablenotetext{c}{Calculated by assuming all the observed \ha\ emission comes from photoionization, Equation (\ref{eq:FHalpha}), with a $\log\left(T/{\rm K}\right)=4.1$.}
\tablenotetext{d}{Compatible properties of a cloud, with $\log\left(n_H/cm^{-3}\right) > -4.0$, that produce the observed \ha\ emission. Quantities are ordered in increasing $n_H$, where $n_H=n_{H_2}+n_{H^0}+n_{H^+}$.}
\tablenotetext{e}{Neutral and ionized line-of-sight depth, where $L_{\rm{H}\textsc{ i}}=N_H/n_H$ and $L_{H^+}=N_e/\left(n_H\times{\rm fraction\ of\ ionized\ H}\right)$. $N_H$ and $N_e$ are the electron and hydrogen column densities and $n_H$ is the total hydrogen number density. }
\tablenotetext{f}{Calculated using the \citet{1999ApJ...510L..33B, 2001ApJ...550L.231B} photoionization model of the Milky Way and the \citet{2001cghr.confE..64H}  model of the extragalactic background radiation. If no value is given, then these model are unsuccessful at  predicting a flux that matches $\log\phi_{LC}(\textrm{\ha})$ any distance.}
\tablenotetext{g}{From absorption-line studies, the distance to core AIII is greater than 8.1~kpc, the distance to core AV is greater than $4.0\pm1.0$~kpc, and the distance to is  core AVI is less than $9.9\pm1.0$~kpc \citep{1996ApJ...473..834W, 1997MNRAS.289..986R, 1999Natur.400..138V, 2003ApJS..146....1W}.  }
\end{deluxetable*}

In general, the Cloudy model produces warm cores with temperatures of $\log\left(T/{\rm K}\right)\sim4.1$. The configuration with well-mixed ionized and neutral gas produces low density cores, electron densities between $-4.0 < \log{\left(n_e/\cm^{-3}\right)} < -1.7$; the scenarios where the ionized gas and neutral gas are either in pressure equilibrium or imbalance are defined by fixing $n_e$ compared to $n_0$. The path length solutions for the neutral gas, determined through observations of the \hi\ gas, are either equivalent to or slightly less than the predicted path length for the ionized gas for the well-mixed and pressure imbalance cases, but greatly differ for the pressure equilibrium configuration. This indicates that the neutral and ionized gas phases are probably well-mixed or in pressure imbalance. However, scenarios where the predicted path length for the ionized gas exceeds a kiloparsec---compared to the smaller path length of the neutral gas that is only a few hundred parsecs long (see Table \ref{table:mass})---could be explained if the Galactic halo provides pressure confinement as hypothesized by \citet{1956ApJ...124...20S}, which would reduce their extent. \citet{1956ApJ...124...20S} shows that a low density, hot, Galactic corona can greatly confine and substantially reduce the path length of a cloud. 

\begin{figure}
\begin{center}
\includegraphics[scale=0.35,angle=0]{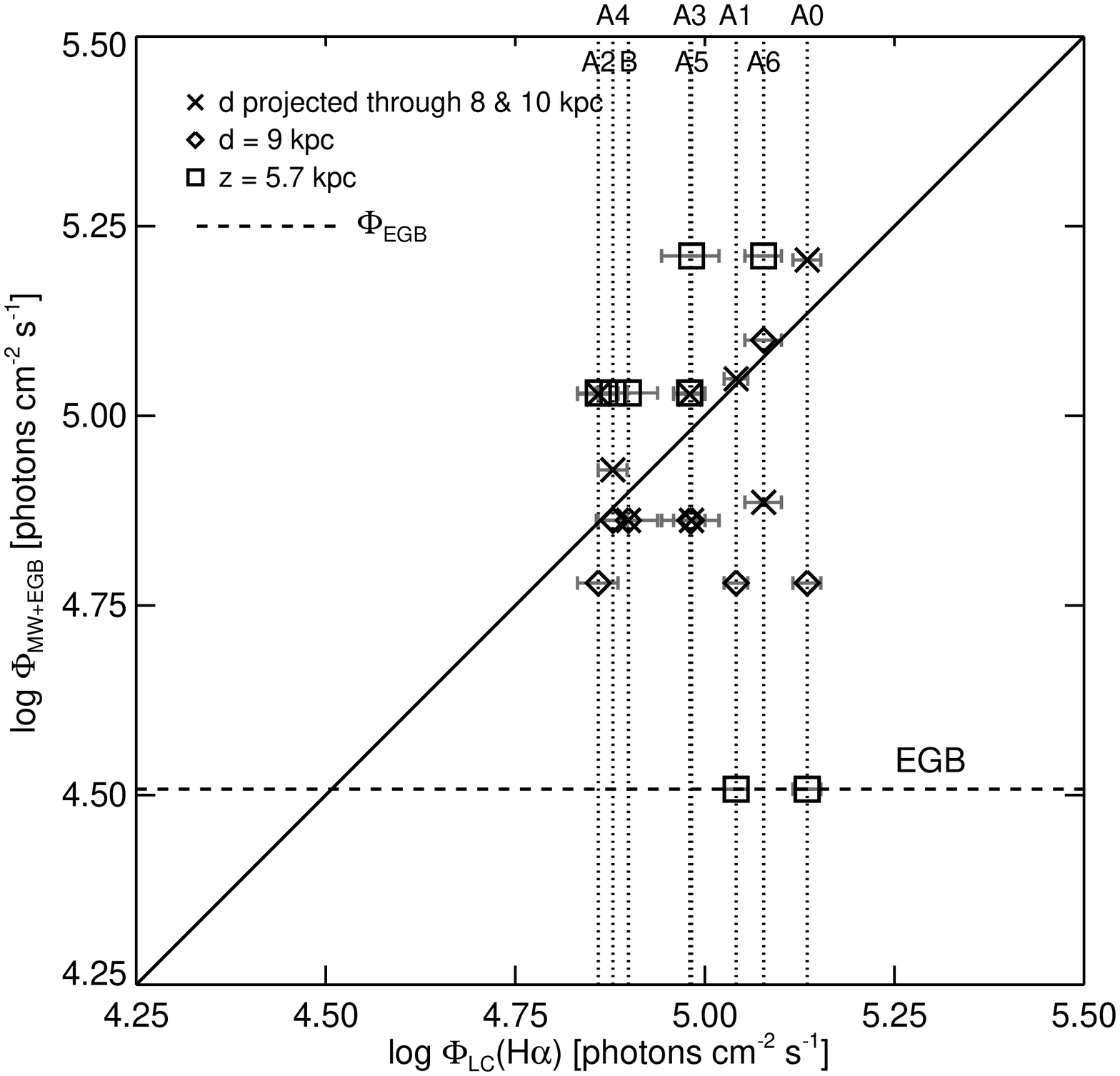} 
\figcaption{
The predicted ionizing flux required to produce the observed \ha\ intensity, $\Phi_{\rm{LC}}(\textrm{\ha})$, compared with the combination $\Phi_{\rm{MW+EGB}}$ of the \citet{1999ApJ...510L..33B, 2001ApJ...550L.231B} and \citet{2001cghr.confE..64H} models, which predict the ionizing flux radiating from the Milky Way and the extragalactic background, $\Phi_{\rm{MW+EGB}}$, respectively. Note that we use the updated version of the Bland-Hawthorn \& Maloney model presented by \citet{2005ApJ...630..332F}. $\Phi_{\rm{LC}}(\textrm{\ha})$ is calculated from Equation (\ref{eq:FHalpha}) for clouds at $\log\left(T/{\rm K}\right) = 4.1$; the corresponding horizontal error bars represent the uncertainty in \ha\ intensity measurement.  $\Phi_{\rm{MW+EGB}}$ shows the fluxes for each core with the same assumed distance ($d=9.0\kpc$), the same height above the galactic plane ($z=5.7\kpc$), and for distances that vary with the allowed projection (see Section \ref{section:distance}). The constant Galactic height orientation is defined by placing the midpoint of cores AIII and AVI at a distance of 9.0 kpc. The solid diagonal line marks the locations where the $\Phi_{\rm{LC}}(\textrm{\ha})$ and the $\Phi_{\rm{MW+EGB}}$ agree.
\label{figure:FHalpha_FJBH}}
\end{center}
\end{figure}

Could the Milky Way and the extragalactic background supply a large enough Lyman continuum flux to produce the observed \ha\ emission? \citet{1999ApJ...510L..33B, 2001ApJ...550L.231B} model the ionizing flux radiating from the Milky Way by assuming that the 90-912 \AA\ radiation is dominated by O-B stars confined to spiral arms\footnote{We use the updated version of the Bland-Hawthorn \& Maloney model presented by \citet{2005ApJ...630..332F} in this study for calculations involving the Lyman continuum flux from the Milky Way.}; this model predicts the shape and strength of both the soft and hard radiation field emitted by the Milky Way. \citet{2001cghr.confE..64H} predict the ionizing flux from the extragalactic background radiation by assuming that the radiation is dominated by quasi-stellar objects, active galaxy nuclei, and active star forming galaxies. We approximate the extragalactic background contribution of ionizing photons as a constant flux with $3.22\times10^4~{\rm photons}\ \cm^{-2}\ \s^{-1}$ along the entire complex. Any attenuation of the extragalactic background flux should be minimal as the cloud is located above the galactic disk. Using the incident ionizing flux from these models, we investigate which distances would reproduce the observed \ha\ emission. Figure \ref{figure:FHalpha_FJBH} illustrates these results for a cloud at a constant distance of 9~\kpc\ from the Sun, for a cloud inclined through (147\fdg7, 35\fdg0) at $8.1~\kpc$ and (160\fdg0, 43\fdg3) at $9.9~\kpc$, and for a cloud at a constant height of $z=5.7~\kpc$ as defined by placing the midpoint of cores~AIII and AVI at a distance of $9.0 \kpc$. The solid diagonal line represents the locations where the predicted ionizing fluxes from the observed \ha\ emission and the Milky Way and extragalactic background flux models match. 

\begin{figure}
\begin{center}
\includegraphics[scale=0.35,angle=0]{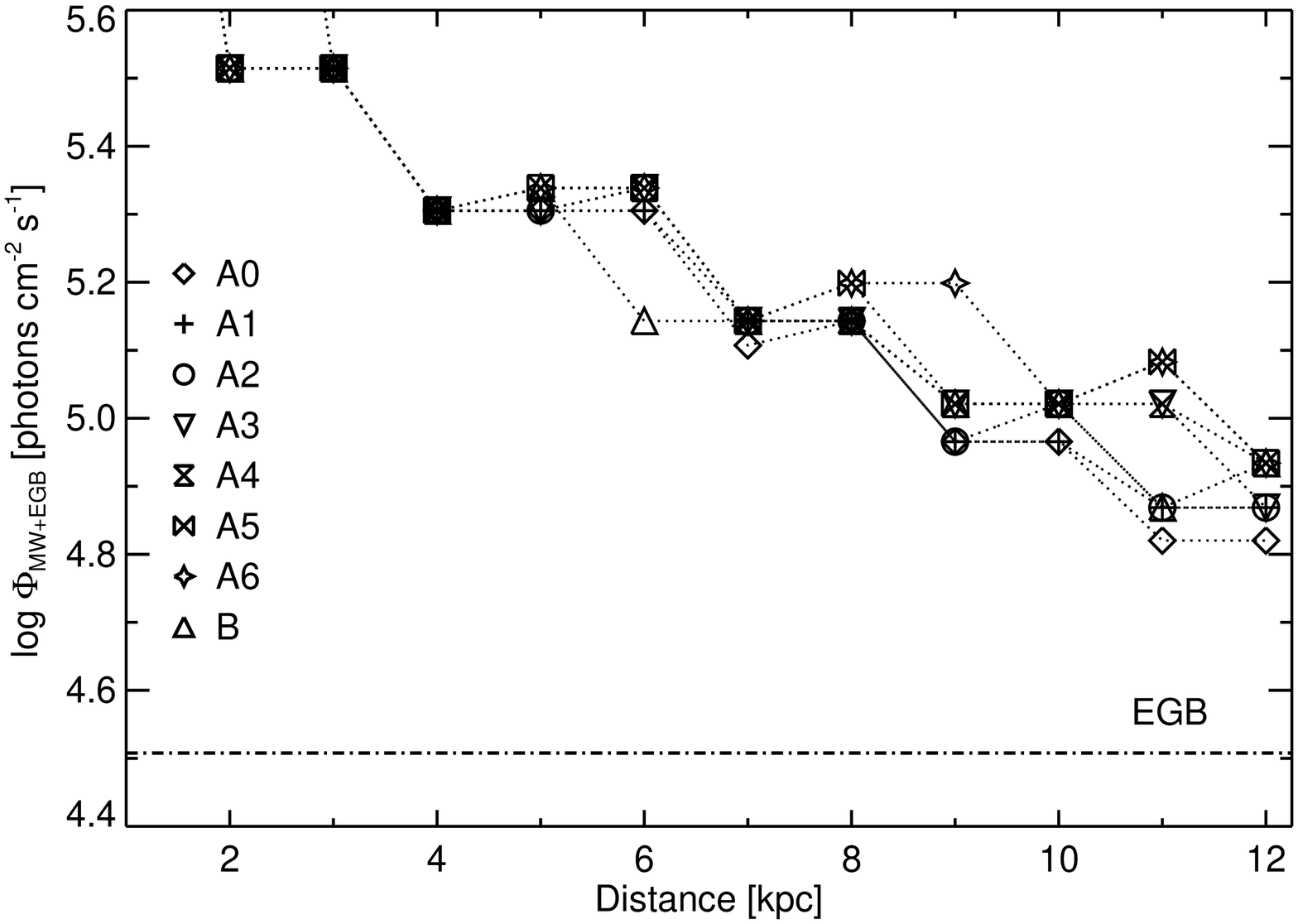} 
\figcaption{
The incident Lyman continuum flux as a function of distance. Using the \citet{1999ApJ...510L..33B, 2001ApJ...550L.231B} model for the ionizing photons escaping from the Milky Way and the \citet{2001cghr.confE..64H} model for the ionizing radiation from the extragalactic background, we estimate the incident ionizing flux in the direction of each core as a function of distance from the Sun. The lowest distance expected for Complex~A is 6.3~kpc (see Section \ref{section:distance} and Table \ref{table:mass}).
\label{figure:Dist_FJBH}}
\end{center}
\end{figure}

The solutions for a Complex~A at constant distance of 9.0~\kpc\ underpredict the ionizing flux for most of the \hi\ cores. The ionizing flux solutions at a constant distance of $z=5.7~\kpc$ underpredict fluxes for cores~A0 and A1 and overpredicts fluxes for the remaining cores. The Lyman continuum solutions for the inclined complex produce a better match, but do not adequately predict the necessary fluxes for every core. Figure \ref{figure:Dist_FJBH} shows that the expected incident Lyman continuum flux from these models varies very little between each core with distance from the Sun. If photoionization from the Milky Way is the dominant ionization source, then the ionizing continuum from both the Milky Way and the extragalactic background can be used to predict the distance that each core would need to reproduce the observed \ha\ emission. Table \ref{table:flux_halpha} lists these distances. These predicted distances agree with the distance constraints found through absorption-line studies; specifically, the distance to core~AIII is greater than $8.1~\kpc$, the distance to core~AV is greater than $4.1\pm1~\kpc$, and the distance to core~AVI is less than $9.9\pm1~\kpc$. 
 
Although the \citet{1999ApJ...510L..33B, 2001ApJ...550L.231B} model predicts enough escaping ionizing radiation from the Milky Way to produce reasonable cloud properties, this model might underpredict the ionizing flux at small distances above the Galactic midplane. \citet{2010ApJ...721.1397W} argue that this model underpredicts the ionizing flux because this model confines all ionizing sources to the Galactic midplane and are attenuated by a uniform density slab of dust. Instead \citet{2010ApJ...721.1397W} suggest that 3D simulations are needed to account for low-density paths and voids that are produced in a turbulent ISM. These voids and low-density paths  allow the ionizing photons from the midplane to penetrate and ionize the Galactic halo. 

The source of the ionization can influence the morphology of Complex~A and of the individual cores. \citet{2000A&A...357..120B} suggest that HVCs interacting with their surrounding ambient medium could have a head-tail morphology, similar to that of a comet, where the tails corresponds to material stripped from the HVC core. Using \hi\ observations from the Leiden/Dwingeloo survey \citep{1997agnh.book.....H}, \citet{2000A&A...357..120B} finds that many of the dense \hi\ cores in Complex~A exhibit this head-tail morphology, as does the Smith Cloud (e.g., \citealt{2008ApJ...679L..21L}). \citet{2007ApJ...670L.109B} finds that as clouds travel through the halo, some of their leading gas can become stripped and trail behind. The trailing gas can then shock the downstream gas, resulting in an ionized skin and fragmentation. The elongated multi-core structure of Complex~A could indicate that this cloud has became ionized and fragmented as it passed through the Galactic halo. The collisionally excited gas will produce variations in the \sii/\nii\ ratio along the length of the cloud \citep{1999ApJ...523..223H}. Unfortunately, the signal-to-noise ratio of the \sii\ and \nii\ lines is too low to constrain the contribution of collisional ionization. Although Lyman continuum fluxes predicted by the \citet{1999ApJ...510L..33B, 2001ApJ...550L.231B} and \citet{2001cghr.confE..64H} models of the ionizing photons originating from the Milky Way and the extragalactic background can produce the observed \ha\ emission and realistic cloud properties, the morphology of Complex~A suggests that interactions with the halo might also contribute. 
 
\section{Electron Temperature Constraints}\label{section:temperature} 

\begin{figure}
\begin{center}
\includegraphics[scale=0.35,angle=90]{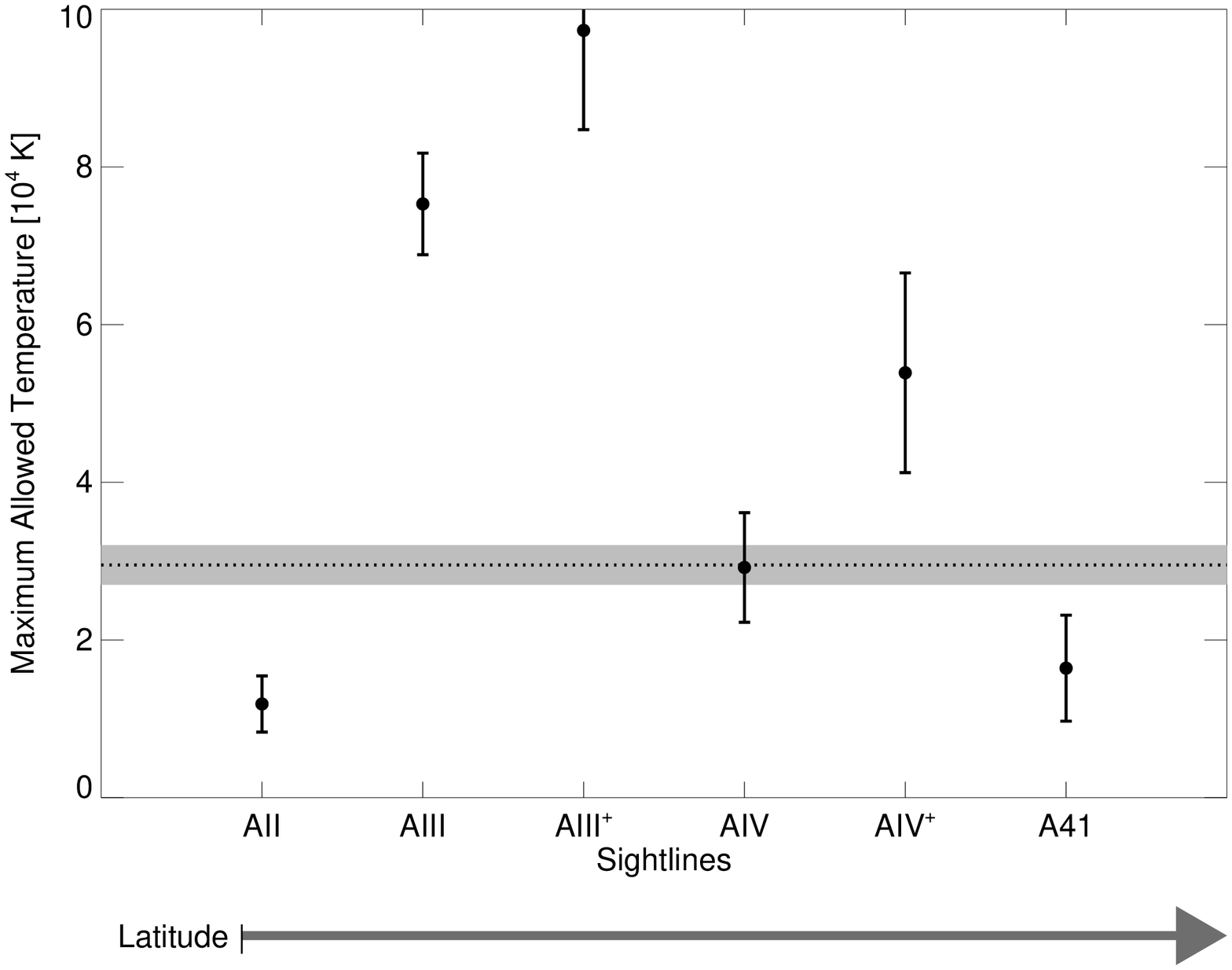} 
\end{center}
\figcaption{
The upper limit of the temperature from the line-widths, calculated using Equation (\ref{eq:RonTemp}) for negligible non-thermal line broadening effects, for the pointed sightlines where the line profile was nearly Gaussian. The horizontal dotted line and corresponding grey region signify the weighted average and error of these points and indicates that the temperature is $\lesssim3.3\times10^4$~K. Note that this temperature limit calculation assumes that the contributions from non-thermal motion are zero. 
\label{figure:Temperature}}
\end{figure}

\citet{1985ApJ...294..256R} derived a method for calculating the temperature of \ha\ emission using its width (FWHM$_{\mha}$) and the velocity profile:
\begin{equation}\label{eq:RonTemp}
\frac{T}{10^4~\mathrm{K}}\approx\left(\frac{\mathrm{FWHM}_{\mha}}{21.4~\mathrm{\kms}}\right)^2-\left(\frac{v_{\mathrm{p,\, non-therm}}}{12.8~\mathrm{\kms}}\right)^2-0.070, 
\end{equation}
where $v_{p,non-therm}$ is the most probable speed (mode) of the non-thermal distribution of gas particles. Ideally we would calculate the $v_{p,non-therm}$ by comparing the width of the \ha\ line with the line width of a much heavier elements like sulfur or nitrogen since bulk motions equally affect gas in equilibrium but thermal motions influence the lighter particles more than the heavy ones. Unfortunately, the low signal-to-noise ratio of the \sii\ and \nii\ observations make measuring their widths difficult. Neglecting non-thermal broadening mechanisms, we calculate the upper limit of the electron temperature due to thermal broadening effects alone. Only a few sightlines exhibit high enough signal-to noise to accurately measure the width of the \ha\ line: AII, AIII, AIV, A$_{41}$, and two pointings just off the centers of cores AIII and AIV. The average, weighted by the error in the $\mathrm{FWHM}_{\mha}^2$, of these solutions gives a temperature of $\left(3.0\pm0.3\right)\times10^4$~K, shown in Figure \ref{figure:Temperature}. This method can only place constraints on the maximum thermal contribution on the line widths and places no constraints on the non-thermal broadening mechanisms.  

The source of the ionization in Complex~A constrains the lower electron temperature limit. In Section \ref{section:ionization}, we found that photoionization can produce the observed \ha\ emission and reasonable cloud properties. If photoionization dominates, then the Cloudy results for the four different inclinations of Complex A included in Table \ref{table:mass} indicate that the lowest reasonable electron temperature consistent with the \ha\ detections is $8000~{\rm K}$. Note that the Cloudy modeling only provides a rough estimate of this lower limit based on the idealized pure photoionized circumstances described in Section \ref{section:ionization}. Figure \ref{figure:FHalpha_AIV} shows that at $8000~{\rm K}$, the expected \ha\ intensity for core AIV drops rapidly for lower electron temperatures. This trend is consistent for each of the cores. 
 
 \section{Total Mass of the Ionized Gas}\label{section:mass}

The Milky Way must be accreting external sources of gas to sustain star formation as its gas consumption time is only 1--2~Gyrs \citep{1980ApJ...237..692L}.  Inflowing low-metallicity gas could supply this material and reproduce the stellar abundance patterns in the Milky Way. HVCs may represent much of this inflow. Quantifying both the neutral and ionized mass in circumgalactic gas structures constrains the accretion rate of galaxies. Four effects dominate the uncertainty in accurately calculating the mass of Complex~A: the distance to the cloud, the projection angle of the cloud, the morphology of the cloud along the line-of-sight, and the distribution of ionized and neutral gas within the cloud. The distance to the cloud, its projection, and the distribution of the ionized and neutral gas are discussed in Section \ref{section:modeling}. When estimating the mass of the cloud, we consider four different cloud projections in Section \ref{section:distance} and three gas distributions in Section \ref{section:distribution}. 

Because the \ha\ emission is proportional to the square of the density, the mass of the ionized gas also depends on the electron density squared. The mass of the ionized gas is given by $M_{H^+}=1.4m_{\rm H} n_e D^2 \Omega L_{H^+}$, where $m_{\rm H}$ is the mass of a hydrogen atom, $\Omega$ is the solid angle, $D$ is the distance to the cloud from the Sun, and the factor of 1.4 accounts for helium. For a  1\arcdeg\ circular beam---the angular resolution of the \ha\ observations---the mass of the ionized gas is given by \citep{2009ApJ...703.1832H}
\begin{equation}
\left({\frac{M_{H^{+}}}{M_\odot}}\right)_{beam}=8.26\left(\frac{D}{\textmd{kpc}}\right)^2 \left(\frac{EM}{\mathrm{pc\cdot cm^{-6}}}\right)\left(\frac{n_e}{\textmd{cm}^{-3}}\right)^{-1}.
\end{equation}
The emission measure, $EM\equiv\int n_e^2\ dL_{H^+}$, is given as  
\begin{equation}\label{eq:EM}
 EM=2.75\ \left(\frac{T}{10^4~{\rm K}}\right)^{0.924} \left(\frac{\iha}{R}\right) \pc \cdot \cm^{-6},    
\end{equation}
using Equation \ref{eq:L_ion} for the $L_{H^+}$ and by assuming that both the electron density and the temperature are constant over the emitting region. Since the more diffuse \ha\ emission generally follows the \hi, we use the neutral core structures to divide the complex into distinct regions where we estimate gas mass. These regions are defined in Table \ref{table:region} and shown in Figures \ref{figure:Map}c--d. Table \ref{table:mass} lists the estimated \hi\ and \ha\ mass for these regions. 

To determine the ionized mass of Complex~A, we exclude the faint \ha\ emission more than 2.5 degrees from the $3\times10^{18}\ \cm^{-2}$ \hi\ column density contours because a definitive association is less certain between the faint \hi\ and \ha\ gas. Therefore, the calculated \ha\ mass should be treated as a minimum. In several of the sightlines that are close to the \hi\ contours, absorption associated with Complex~A is absent; this is true for both highly-ionized ions, such as O{\sc~vi}, and for low-ionization species, such as C{\sc~ii} \citep{2006ApJS..165..229F}. As a result, the extended ionized envelope around Complex~A may be patchy.

The largest uncertainty in determining the mass of Complex~A comes from the uncertainty in distance for each of the high \hi\ column density cores. Section \ref{section:distance} discusses the four projections used to estimate the distance to each of these regions. The $\rm{H}^0$ mass estimate for Complex~A ranges from $1.6$ to $2.3\times10^6~M_{\odot}$ depending on the projection used. The $\rm{H}^+$ mass estimate has a larger variation due to the unknown distribution of the ionized gas.  One of these distributions, where the skin is in pressure equilibrium with the \hi\ such that $n_e=1/2n_0$, seems unphysical as the solutions give line-of-sight distances on the order of a few kiloparsecs. The $\rm{H}^+$ mass ranges from $1.3$ to $2.5\times10^6~M_{\odot}$ depending on the projection angle and the distribution of gas---excluding the distribution where the \ha\ skin is in pressure equilibrium with its surroundings. From these estimates, the mass of the $\rm{H}^+$ gas appears to be at least comparable to the neutral gas of Complex~A. \\

\section{Sulfur \& Nitrogen Abundance}\label{section:abundance}

\begin{figure}
\begin{center}
%\plottwo{figures/S_H_T_AIV.eps}{figures/S_H_T_Mrk116.eps} 
\includegraphics[scale=.3,angle=90]{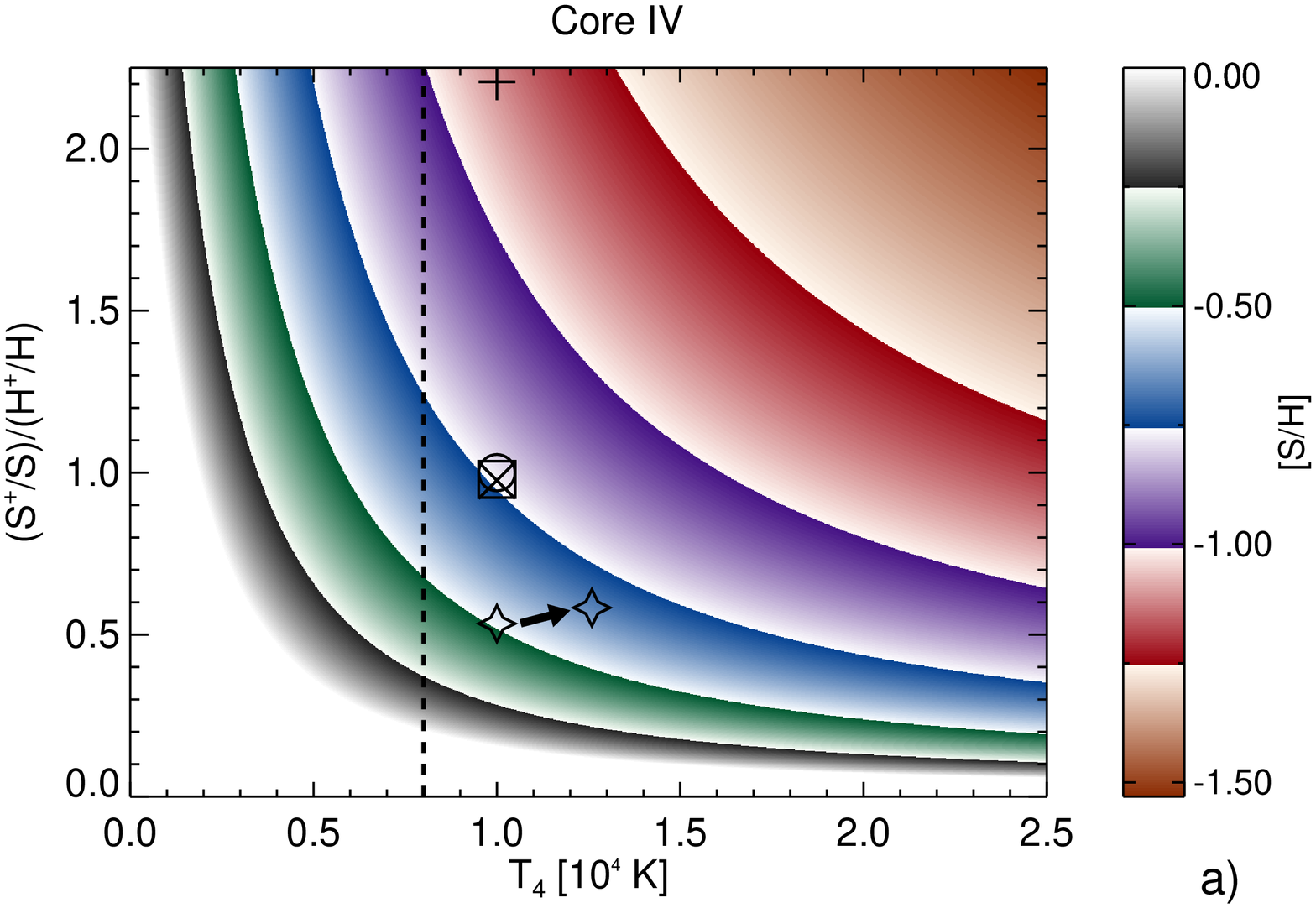}
\includegraphics[scale=.3,angle=90]{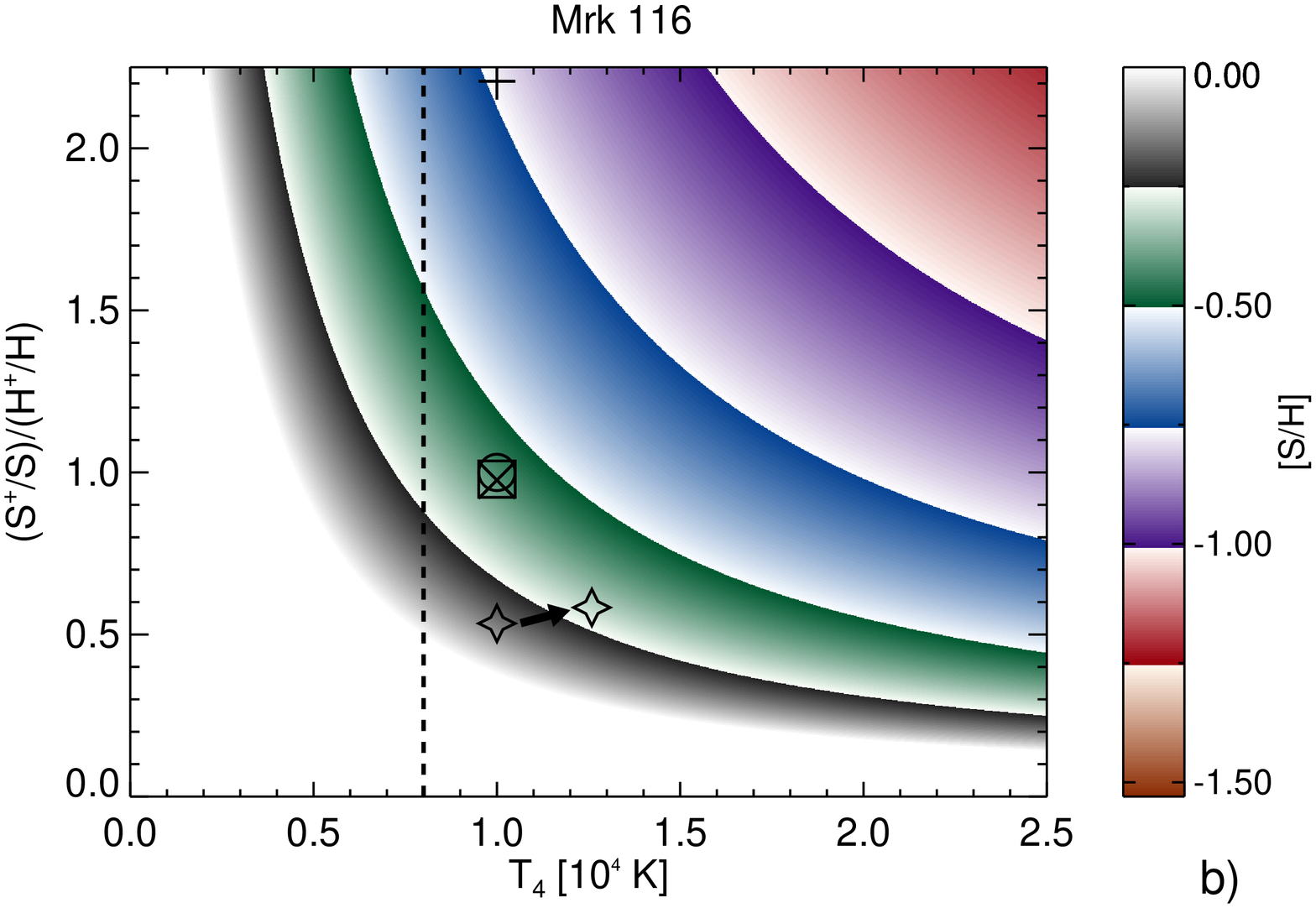} 
\end{center}
\figcaption{The sulfur abundance as a function of temperature and ionization fraction---see Equation (\ref{eq:siiha})---for two of the \hi\ sightlines: \emph{(a)} Core~AIV$^\dagger$ for $I_{[\rm S\textsc{~ii}]}/I_{\rm H\alpha}=0.32\pm0.23$ and \emph{(b)} Mrk~116 for $I_{[\rm S\textsc{~ii}]}/I_{\rm H\alpha}=0.78\pm0.59$. The sulfur abundance (${\rm S}/\H \equiv S_{total}/\H_{total}$; $[{\rm S}/\H] \equiv \log({\rm S}/\H) - \log({\rm S}/\H)_{\odot}$) and $({\rm S}/ \H)_{\odot} = 1.3 \times 10^{-5}$; \citealt{2009ARA&A..47..481A}) is derived from the observed intensity ratio from the values listed in Table \ref{table:intensity_time}. The overlaid symbols mark the metallicity solutions for different cloud configurations, including different projections (see Table \ref{table:metallicity}). The vertical dashed line at $8000~{\rm K}$ denotes the typical photoionization temperature found in the warm ionized medium of the Milky Way \citep{1998ApJ...501L..83H, 1999ApJ...523..223H}; temperatures above $8000~{\rm K}$ are consistent for a purely photoionized Complex~A as discussed in Section \ref{section:implications}.
\label{figure:S_H_T_CoreIV_Mrk116}}
\end{figure}

\begin{figure}
\begin{center}
%\plottwo{figures/N_H_T_CoreIII.eps} {figures/N_H_T_Mrk106.eps} 
\includegraphics[scale=.3,angle=90]{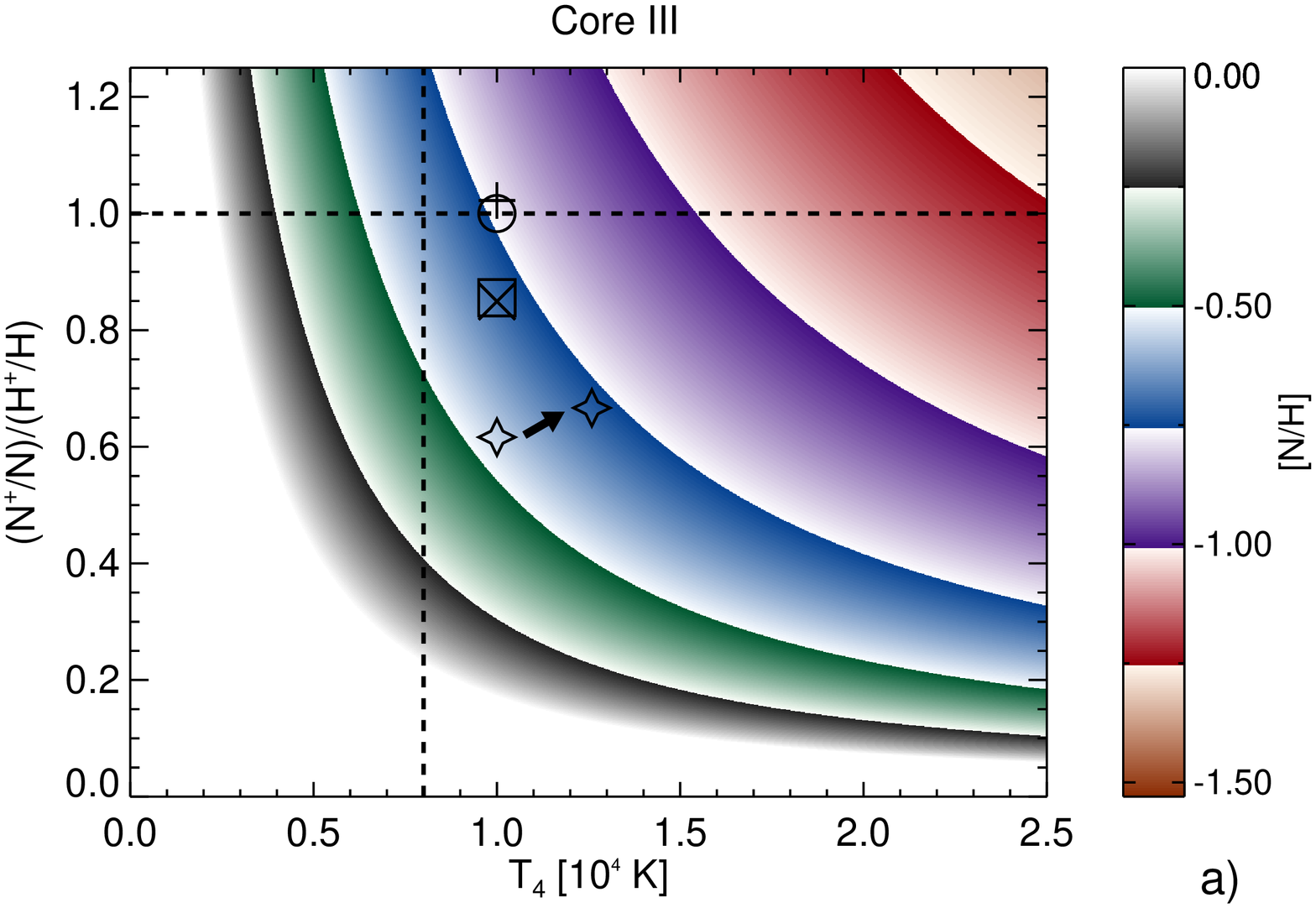} 
\includegraphics[scale=.3,angle=90]{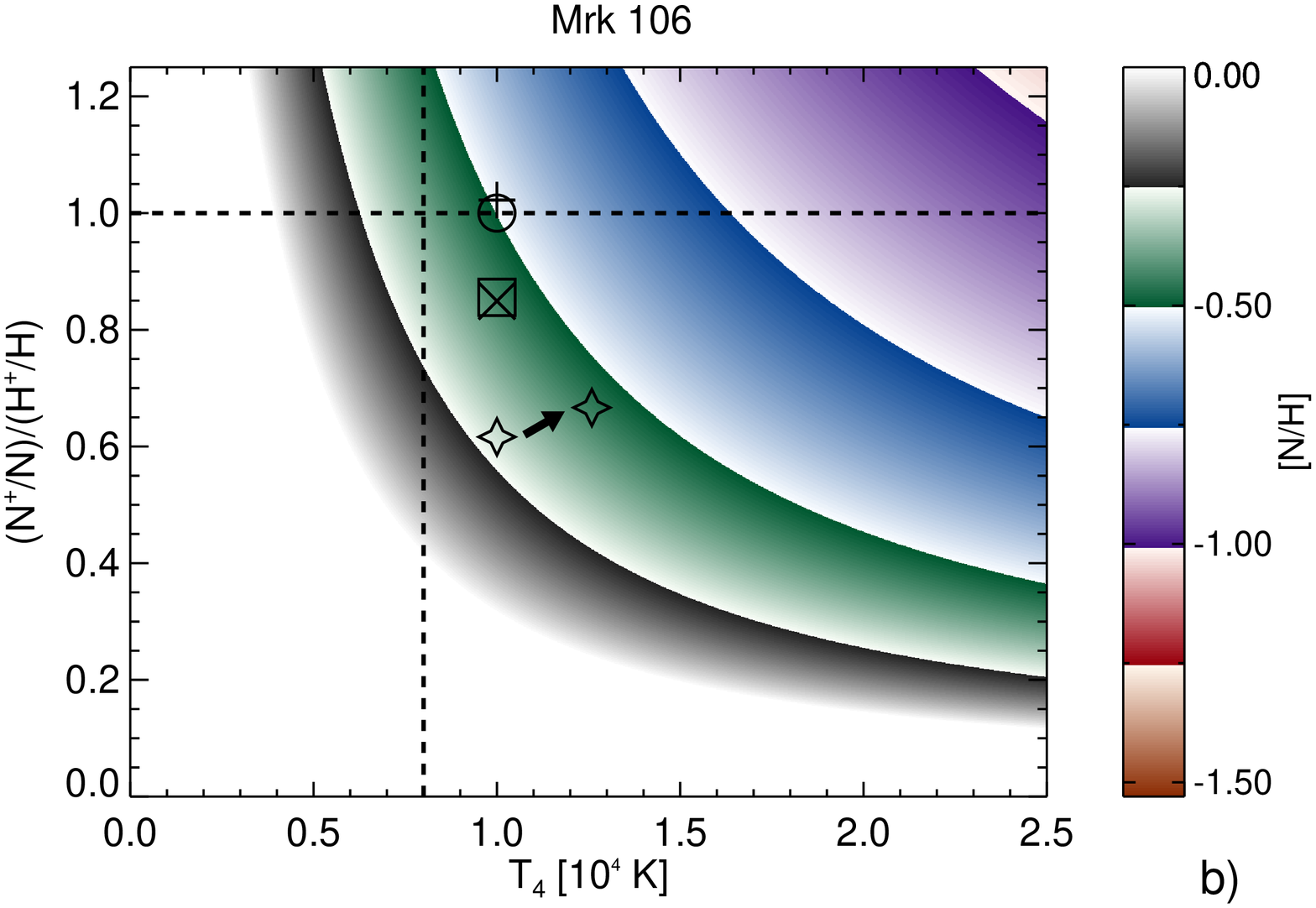} 
\figcaption{The nitrogen abundance as a function of temperature and ionization fraction---see Equation (\ref{eq:siiha})---for two of the \hi\ sightlines: \emph{(a)} Core~AIII for $I_{[\rm N\textsc{~ii}]}/I_{\rm H\alpha}=0.38\pm0.14$ and \emph{(b)} Mrk~106 for $I_{[\rm N\textsc{~ii}]}/I_{\rm H\alpha}=0.85\pm0.14$. The nitrogen abundance ($\N/\H \equiv \N_{total}/\H_{total}$; $[\N/\H] \equiv \log(\N/\H) - \log(\N/\H)_{\odot}$  and $(\N / \H)_{\odot} = 6.8 \times 10^{-5}$; \citealt{2009ARA&A..47..481A}) is derived from the observed intensity ratio from the values listed in Table \ref{table:intensity_time}. The overlaid symbols mark the metallicity solutions for different cloud configurations, including different projections (see Table \ref{table:metallicity}). The vertical dashed line at $8000~{\rm K}$ denotes the typical photoionization temperature found in the warm ionized medium of the Milky Way \citep{1998ApJ...501L..83H, 1999ApJ...523..223H}; temperatures above $8000~{\rm K}$ are consistent for a purely photoionized Complex~A as discussed in Section \ref{section:implications}. The horizontal line signifies the expected nitrogen and hydrogen ionization fraction ratio for photoionized gas undergoing typical ionization conditions with typical gas properties \citep{2000ApJ...528..310S}; $N^+/N\sim H^+/H$ when the hydrogen and nitrogen charge exchange process is significant, but this ratio could differ for extremely low density gas that is ionized by a high incident flux of ionizing photons.  
\label{figure:N_H_T_CoreIII_Mrk106}}
\end{center}
\end{figure}

Determining the metallicity of Complex~A limits the possible formation mechanisms that created this cloud  and provides information on the material the Milky Way accretes. Four sightlines along the length of Complex~A constrain the composition of the cloud: The sightlines toward Mrk~106 and core~AIII constrain the nitrogen abundance; the sightlines toward  Mrk~116 and core~AIV restrict the sulfur abundance. These four sightlines are the only directions with $> 3 \sigma$ detections of emission in metal lines.

Hydrogen recombination produces \ha\ emission and collisional excitation produces \nii $\lambda6583$ and \sii $\lambda6716$ emission lines; these interactions make these three spectral lines sensitive to the electron temperature, the ionization fraction of the gas, and the electron density. The intensity ratios eliminate the geometry dependency by removing the electron density---assuming the emission originates from the same region. The \nii/\ha\ intensity ratio towards the same sightline is given by \citep{1999ApJ...523..223H}
\begin{equation} \label{eq:niiha}
\frac{\mnii}{\mha} = 1.63 \times 10^5 \left[\left(\frac{\Hplus}{\H}\right)^{-1}   \frac{\N^+}{\N} \right] \left(\frac{\N}{\H}\right) \left(\frac{T}{10^4~{\rm K}}\right)^{0.426} e^{\left[-2.18/\left(T/10^4~{\rm K}\right)\right]}, 
\end{equation}
where N$^+$/N and H$^+$/H are the ionization fractions and $\mathrm{N} / \mathrm{H}$ is the gas phase abundance by number.
Similarly, the \sii/\ha\ intensity ratio is given by
\begin{equation} \label{eq:siiha}
\frac{\msii}{\mha} = 7.64 \times 10^5 \left(\frac{\Hplus}{\H}\right)^{-1} \left(  \frac{\rm{S}^+}{\rm{S}} \frac{\rm{S}}{\H} \right) \left(\frac{T}{10^4~{\rm K}}\right)^{0.307} e^{\left[-2.14/\left(T/10^4~{\rm K}\right)\right]}. 
\end{equation}
where S$^+$/S is the ionization fraction and $\mathrm{S} / \mathrm{H}$ is the gas phase abundance by number.

\begin{deluxetable*}{lcccccccccccccc}
\tabletypesize{\scriptsize}
\tablecaption{Properties of Photoionized Regions with $\phi_{LC}(\textrm{\ha})$ estimated from \ha\ Observations\label{table:metallicity}} 
\tablewidth{0pt}
\tablehead{
\multicolumn{2}{c}{}	& \colhead{} & 	\colhead{}& \multicolumn{3}{c}{Ionization Fraction\tablenotemark{a}}	& \colhead{} & \multicolumn{2}{c}{$\left(N/H\right)/\left(N/H\right)_{\odot}$} & \colhead{} & \multicolumn{2}{c}{$\left(S/H\right)/\left(S/H\right)_{\odot}$} \\
\cline{5-7} \cline{9-10} \cline{12-13}\\
\colhead{Symbol\tablenotemark{b}}	& \colhead{Configuration}	& \colhead{$\log{\rm T/\rm K}$\tablenotemark{c}} & \colhead{$\log{n_e/\cm^{-3}}$\tablenotemark{c}} & \colhead{$N^+/N$}	 &\colhead{$S^+/S$}	& \colhead{$H^+/H$}	& \colhead{} & \colhead{AIII\tablenotemark{d}} & \colhead{Mrk~116\tablenotemark{d}} 	& \colhead{} 				& \colhead{AIV\tablenotemark{d}}		& \colhead{Mrk~106\tablenotemark{d}} }
\startdata				  
$\diamondsuit$		& $\Phi(\mha, 10^4~{\rm K})$\tablenotemark{e}		& 4.1 & -2.4 & 0.45 	& 0.39 & 0.73			&& $0.51\pm0.18$ 	& $1.10\pm0.19$	&& $0.51\pm0.38$	& $1.23\pm0.93$	\\
$\diamondsuit$		& $\Phi(\mha, 10^{4.1}~{\rm K})$\tablenotemark{e}	& 4.1 & -2.4 & 0.48	& 0.42 & 0.72			&& $0.27\pm0.19$	& $0.59\pm0.10$	&& $0.28\pm0.21$	& $0.68\pm0.51$	\\
$+$				& $d=8-10~{\rm kpc}$						& 3.9 & -1.5 & 0.37	& 0.79 & 0.36			&& $0.30\pm0.11$	& $0.66\pm0.11$	&& $0.12\pm0.09$	& $0.30\pm0.23$	\\
$\square$			& $d=9~{\rm kpc}$							& 4.0 & -2.0 & 0.50	& 0.57 & 0.58			&& $0.36\pm0.13$	& $0.79\pm0.13$	&& $0.28\pm0.20$	& $0.67\pm0.51$	\\
$\times$			& $z=5.7~{\rm kpc}$							& 4.0 & -2.1 & 0.50	& 0.50 & 0.58			&& $0.37\pm0.13$	& $0.80\pm0.14$	&& $0.28\pm0.21$	& $0.67\pm0.51$	\\
o				& $N^+/N=S^+/S=H^+/H$\tablenotemark{f}		& --    & -- & $H^+/H$& $H^+/H$ & $H^+/H$	&& $0.37\pm0.13$	& $0.80\pm0.14$	&& $0.28\pm0.21$	& $0.68\pm0.51$	 \\
\enddata
\tablenotetext{a}{Values based on Cloudy photoionization modeling of core~AIV for each configuration, except for when $N^+/N=S^+/S=H^+/H$.}
\tablenotetext{b}{Symbols used in Figures \ref{figure:S_H_T_CoreIV_Mrk116} \& \ref{figure:N_H_T_CoreIII_Mrk106}.}
\tablenotetext{c}{The average $\log{\rm T/\rm K}$ and $\log{n_e/\cm^{-3}}$ that produce the observed \ha\ emission, predicted through Cloudy modeling. }
\tablenotetext{d}{$I_{\rm S{\textsc{~ii}}}/{I_{\mha}}$ ratio for AIV is $0.32\pm0.23$ and $0.78\pm0.59$ for Mrk~116. $I_{\rm N{\textsc{~ii}}}/{I_{\mha}}$ ratio for AIII is $0.38\pm0.14$ and $0.85\pm0.14$ for Mrk~106. }
\tablenotetext{e}{Assumes that the incident ionizing flux is proportional to the \ha\ intensity using Equation \ref{eq:FHalpha}.}
\tablenotetext{f}{If the nitrogen ionization fraction is governed by charge exchange reactions with hydrogen, then $N^+/N\sim H^+/H$; this assumption is typically used when investigating the WIM 
since the hydrogen and nitrogen first ionization potential are similar (N: 14.4~eV and H: 13.6~eV). $S^+/N\sim H^+/H$ is listed for symmetry. These solutions are determined from Equations (\ref{eq:niiha}) and (\ref{eq:siiha}).}
\end{deluxetable*}

When the ionization fractions of the singly ionized species and the gas temperature are known, Equations (\ref{eq:niiha}) and (\ref{eq:siiha}) can be used in conjunction with our observations to constrain the gas-phase abundances. Figures \ref{figure:S_H_T_CoreIV_Mrk116} and \ref{figure:N_H_T_CoreIII_Mrk106} show this relationship for a reasonable range of  ionization fractions and temperatures constrained by our best-determined ratios: \sii/\ha\ toward core~AIV$^\dagger$ and Mrk~116 and \nii/\ha\ toward core~AIII and Mrk~106. Symbols in these figures denote estimates for the ionization fractions at a fixed temperature and configuration and are summarized in Table~\ref{table:metallicity}. Cloudy modeling of core~AIV, using the input geometry discussed in Section~\ref{section:distance}, determines five of these values. The gas temperature in the emitting region is assumed to be $10^4~{\rm K}$ in all cases except one where we use $10^{4.1}~{\rm K}$ reflecting the discussion in Section \ref{section:ionization}. We also show a solution where the singly-ionized nitrogen, sulfur, and hydrogen ionization fractions are equal. While $\mathrm{N}^+/\mathrm{N} \sim \mathrm{H}^+/\mathrm{H}$ is a reasonable assumption for diffusely ionized gas with similar ionization potentials and a weak charge-exchange reaction, this assumption is likely an over-simplification for sulfur as this ratio can vary in warm ionized gas due of the low second ionization potential $(23.4~{\rm eV})$ of sulfur (e.g., \citealt{1999ApJ...523..223H, 2006ApJ...652..401M}). 

As Figures \ref{figure:S_H_T_CoreIV_Mrk116} and \ref{figure:N_H_T_CoreIII_Mrk106} show, without tighter constraints on the ionization fraction and temperature, this current dataset does not place tight constraints on the nitrogen and sulfur abundances. Previous studies suggest that this cloud has an abundance of roughly a tenth solar (e.g., \citealt{1994A&A...282..709K, 1995A&A...302..364S, 1996ApJ...473..834W, 1999Natur.400..138V, 2001ApJS..136..537W}). The abundance solutions, determined with the observed line ratios, become consistent with previous studies when the gas has a temperature above$10^4~{\rm K}$ or when the gas has a high singly-ionized fraction of nitrogen and sulfur (see Table \ref{table:metallicity} and Figures \ref{figure:S_H_T_CoreIV_Mrk116} and \ref{figure:N_H_T_CoreIII_Mrk106}).

\section{Implications of the Ionized Gas}\label{section:implications}

The metallicity of HVCs provides the one of the tightest constraints on possible origins of the cloud. This study finds that many of the possible cloud configurations have a sub-solar metallicity, which drops rapidly with electron temperature (see Figures \ref{figure:S_H_T_CoreIV_Mrk116} and \ref{figure:N_H_T_CoreIII_Mrk106}). The Cloudy modeling of a cloud bathed in Lyman Continuum determined by the \ha\ emission suggests that the electron temperature is above $8000~{\rm K}$ to be consistent with the measured \ha\ intensities. The \ha\ line widths of the bright targeted observations suggest that the electron temperature is less than $3.3\times10^4\ \rm{K}$, as determined from the weighted average of the temperature determined for six sightlines. Many of the cloud configurations place the Complex~A abundances below solar, which may indicate that the cloud originated from somewhere other than the Milky Way disk. 

Other studies also suggest a sub-solar composition for Complex~A (e.g., the oxygen abundance is between 0.03 and 0.4 solar:  \citealt{1994A&A...282..709K} \& \citealt{2001ApJS..136..537W}; the magnesium abundance is greater than 0.035 solar: \citealt{1996ApJ...473..834W}; $N_{\textrm{Ca{\sc~ii}}}/\textrm{\nhi} = 19\times10^{-9}\pm5\times10^{-9}$: \citealt{1995A&A...302..364S} \& \citealt{1999Natur.400..138V}) which indicates that the cloud probably originated from material stripped off a nearby satellite galaxy or from low-metallicity intergalactic gas; however, no complementary stellar stream has been found to suggest that this complex has been stripped from a dwarf galaxy. Complex A does lie in the projected orbital path of the Orphan Steam \citep{2007ApJ...658..337B}, but the kinematics of are not consistent with an association \citep{2010ApJ...711...32N}. Though this and previous studies investigate the metallicity of Complex~A, neither tightly constrain its metallicity and further studies are needed to constrain the origin of this cloud. 

The ionized gas is a major component of the total mass in Complex~A. The incident radiation from the Milky Way and the extragalactic background can produce the observed \ha\ emission. However, the long multi-core morphology of the Complex~A suggest that interactions with the halo also contribute. The morphology could indicate that the outer layers of the cores are being stripped and are shocking the trailing gas; this could cause the cloud to fragment while traveling through the halo \citep{2009IAUS..254..241B}. The shock model presented by \citet{1979ApJS...39....1R} predicts the production of range of low-ionization emission lines, such as \sii\ and \oi; the signal-to-noise ratios of the observations of \sii\ and \oi\ are too low to identify shocks as a major ionization source of these lines. The two sightlines with detected \sii\ show a large variation in \sii/\ha, possibly indicating that these two regions experience large variations in ionization or excitation conditions \citep{1999ASPC..168..149R}. Furthermore, a shock excitation scenario could explain the non-uniform \ha\ intensities along the length of the cloud as the \ha\ flux would vary strongly with the shock and ambient density \citep{1998ApJ...504..773T}. 

Though the distance to Complex~A has been constrained to be about 8 to 10~kpc towards one foreground star and one background star (\citealt{1997MNRAS.289..986R} \citealt{1999Natur.400..138V}, and \citealt{2003ApJS..146....1W}), the morphology of the complex makes its projection uncertain and creates a large ambiguity in determining the total mass of the cloud. To anchor the projection angle, more locations along the length of the cloud need their distances measured. This discrepancy must be resolved to understand how much replenishment this cloud offers the Milky Way and to understand how this cloud interacts with the Galactic halo, since the density of the halo varies with distance from the Galaxy.

Populations of mostly neutral, partially ionized, and highly ionized HVCs exist around other galaxies (e.g., M~101, NGC~891, NGC~2403, NGC~6946, M~83, M~33; see \citealt{2008A&ARv..15..189S} for a review), but little is known about the extent of the $10^4~\rm{K}$ ionized gas. Though many studies have investigated the hot $(10^6~\rm{K})$ component of HVCs, these studies require background sources to study this highly ionized medium; this makes quantifying the extent of the hot gas difficult. Further investigations of the ionized component of these clouds will help discern the extent and the source of this ionized gas. Understanding all the gas phases in HVCs is essential to unraveling how these clouds affect galaxy evolution.

\section{Summary}\label{section:summary}

Using WHAM to observe the warm gas phase in Complex~A, we mapped the \ha\ emission over $1350~{\rm degrees}^2$. These kinematically resolved observations---over the velocity range of -250 to -50~\kms\ in the local standard of rest reference frame---include the first full \ha\ intensity map of Complex~A. We compare these observations of the warm ionized gas phase with the \hi\ $21~\cm$ emission in the LAB survey. Additionally, we include 31 deep targeted observations in \ha\ and 8 in \sii $\lambda6716$, \nii $\lambda6584$, and \oi $\lambda6300$ towards \hi\ dense regions, background quasars, and stars to investigate the physical properties of this cloud. This study finishes with five main conclusions from the observations and Cloudy modeling of the warm gas component in Complex~A:

\begin{enumerate}

\item {\bf Ionizing Sources.} The properties of a purely photoionized cloud, modeled with Cloudy, can reproduce clouds with reasonable properties (see Table \ref{table:flux_halpha}). For a purely photoionized cloud, the \citet{1999ApJ...510L..33B, 2001ApJ...550L.231B} and \citet{2001cghr.confE..64H} Lyman continuum models of the Milky Way and extragalactic background, respectively, place the distances to dense \hi\ cores between 8.1 and 11.5~\kpc, just within the distance bounds inferred through absorption-line studies.

\item {\bf Ionized Skin.} The \ha\ morphology follows the global \hi\ distribution, but the \ha\ emission does not closely trace the \hi\ emission on small scales (see Figure \ref{figure:Map}). At the \hi\ cores, the \ha\ emission often has a lower intensity and an offset velocity distribution. This indicates that the \ha\ and \hi\ emission trace different paths through the clouds---warm ionized gas probably surrounds cold cores. This skin is likely not in pressure equilibrium with the \hi\ because that equilibrium would result in extreme line-of-sight lengths for the ionized gas. 

\item{\bf Ionized Gas Mass.} The total mass of the ionized gas in Complex~A depends on four still uncertain factors: the distance to the cloud, the projection angle of the cloud, the morphology of the cloud along the line-of-sight, and the distribution of the neutral and ionized gases within the cloud. The distance to the high latitude end of Complex~A is between $8$ and $10~\kpc$; however, the distances to the lower latitude \hi\ cores are not yet constrained by direct distance measurements. The distribution of the ionized gas either envelops the neutral gas---where the ionized gas is not in pressure equilibrium with the \hi---or partially mixes with the neutral gas. Accounting for each of these factors, the total mass of the ionized gas could range from $1.3$ to $2.5\times10^6~M_{\odot}$ compared to $1.6$ to $2.3\times10^6~M_{\odot}$ for the neutral component; these mass estimates exclude the extended \ha\ emission that is more than 2.5 degrees off the $3\times10^{18}\,\cm^{-2}$ \hi\ column density contour. As a result, the ionized component makes up a significant fraction of the overall composition of Complex A.

\item {\bf Temperature of the Ionized Component.} The \ha\ line width constrains the maximum kinetic electron temperature of the gas. Comparing \ha\ line widths toward sightlines with bright emission, we find an upper limit of $10^{4.5}~{\rm K}$ for this temperature. %Additionally, the source of the ionization constrains the lower electron temperature limit. If photoionization is the dominant ionization mechanism, then the lowest reasonable electron temperature consistent with the \ha\ detections is $8000 {\rm K}$ as predicted through Cloudy modeling and as shown in Figure \ref{figure:FHalpha_AIV}.     

\item {\bf Sulfur \& Nitrogen Abundance.} The sulfur and the nitrogen detections towards four sightlines reveal a sub-solar composition over a wide range of possible electron temperatures and ionization fractions, as shown in Figures \ref{figure:S_H_T_CoreIV_Mrk116} and \ref{figure:N_H_T_CoreIII_Mrk106}; however, the \sii\ and \nii\ detections presented in this study do not place tight constraints on the abundance. These figures illustrate that different assumed projection angles of Complex~A cause variations in the metallicity solution. These solutions combined with previous abundance studies of Complex~A suggests that many configurations are only plausible for gas above $10^{4}~{\rm K}$ or with a high singly-ionized fraction of nitrogen and sulfur.

\end{enumerate}
 
\acknowledgments The authors acknowledge helpful discussions with Bob Benjamin and Joss Bland-Hawthorn and the anonymous referee for comments and suggestions, which greatly improved the paper. We thank Joss Bland-Hawthorn for providing us with the model for the ionizing radiation emitted from the Milky Way and Kurt Jaehnig for his outstanding technical support in the continuing operation of the WHAM instrument. The National Science Foundation supported WHAM  through grants AST~0204973, AST~0607512, and AST~1108911. A. K. D. received support through REU site award AST-0453442. The infrastructure at Kitt Peak is made possible by National Optical Astronomy Observatory and the Tohono OÕodham Nation. Kitt Peak National Observatory, National Optical Astronomy Observatory, which is operated by the Association of Universities for Research in Astronomy (AURA) under cooperative agreement with the National Science Foundation.

{\it Facility:} \facility{WHAM}
\bibliographystyle{apj} 
\bibliography{References}

\begin{thebibliography}{72}
\expandafter\ifx\csname natexlab\endcsname\relax\def\natexlab#1{#1}\fi

\bibitem[{{Asplund} {et~al.}(2009){Asplund}, {Grevesse}, {Sauval}, \&
  {Scott}}]{2009ARA&A..47..481A}
{Asplund}, M., {Grevesse}, N., {Sauval}, A.~J., \& {Scott}, P. 2009, \araa, 47,
  481

\bibitem[{{Belokurov} {et~al.}(2007){Belokurov}, {Evans}, {Irwin},
  {Lynden-Bell}, {Yanny}, {Vidrih}, {Gilmore}, {Seabroke}, {Zucker},
  {Wilkinson}, {Hewett}, {Bramich}, {Fellhauer}, {Newberg}, {Wyse}, {Beers},
  {Bell}, {Barentine}, {Brinkmann}, {Cole}, {Pan}, \&
  {York}}]{2007ApJ...658..337B}
{Belokurov}, V., {Evans}, N.~W., {Irwin}, M.~J., {et~al.} 2007, \apj, 658, 337

\bibitem[{{Besla} {et~al.}(2010){Besla}, {Kallivayalil}, {Hernquist}, {van der
  Marel}, {Cox}, \& {Kere{\v s}}}]{2010ApJ...721L..97B}
{Besla}, G., {Kallivayalil}, N., {Hernquist}, L., {et~al.} 2010, \apjl, 721,
  L97

\bibitem[{{Bland-Hawthorn}(2009)}]{2009IAUS..254..241B}
{Bland-Hawthorn}, J. 2009, in IAU Symposium, Vol. 254, IAU Symposium, ed.
  {J.~Andersen, J.~Bland-Hawthorn, \& B.~Nordstr{\"o}m}, 241--254

\bibitem[{{Bland-Hawthorn} \& {Maloney}(1999)}]{1999ApJ...510L..33B}
{Bland-Hawthorn}, J., \& {Maloney}, P.~R. 1999, \apjl, 510, L33

\bibitem[{{Bland-Hawthorn} \& {Maloney}(2001)}]{2001ApJ...550L.231B}
---. 2001, \apjl, 550, L231

\bibitem[{{Bland-Hawthorn} {et~al.}(2007){Bland-Hawthorn}, {Sutherland},
  {Agertz}, \& {Moore}}]{2007ApJ...670L.109B}
{Bland-Hawthorn}, J., {Sutherland}, R., {Agertz}, O., \& {Moore}, B. 2007,
  \apjl, 670, L109

\bibitem[{{Bland-Hawthorn} {et~al.}(1998){Bland-Hawthorn}, {Veilleux}, {Cecil},
  {Putman}, {Gibson}, \& {Maloney}}]{1998MNRAS.299..611B}
{Bland-Hawthorn}, J., {Veilleux}, S., {Cecil}, G.~N., {et~al.} 1998, \mnras,
  299, 611

\bibitem[{{Bregman}(1980)}]{1980ApJ...236..577B}
{Bregman}, J.~N. 1980, \apj, 236, 577

\bibitem[{{Br{\"u}ns} {et~al.}(2000){Br{\"u}ns}, {Kerp}, {Kalberla}, \&
  {Mebold}}]{2000A&A...357..120B}
{Br{\"u}ns}, C., {Kerp}, J., {Kalberla}, P.~M.~W., \& {Mebold}, U. 2000, \aap,
  357, 120

\bibitem[{{Cardelli} {et~al.}(1989){Cardelli}, {Clayton}, \&
  {Mathis}}]{1989ApJ...345..245C}
{Cardelli}, J.~A., {Clayton}, G.~C., \& {Mathis}, J.~S. 1989, \apj, 345, 245

\bibitem[{{Diplas} \& {Savage}(1994)}]{1994ApJ...427..274D}
{Diplas}, A., \& {Savage}, B.~D. 1994, \apj, 427, 274

\bibitem[{{Ferland} {et~al.}(1998){Ferland}, {Korista}, {Verner}, {Ferguson},
  {Kingdon}, \& {Verner}}]{1998PASP..110..761F}
{Ferland}, G.~J., {Korista}, K.~T., {Verner}, D.~A., {et~al.} 1998, \pasp, 110,
  761

\bibitem[{{Fox} {et~al.}(2006){Fox}, {Savage}, \&
  {Wakker}}]{2006ApJS..165..229F}
{Fox}, A.~J., {Savage}, B.~D., \& {Wakker}, B.~P. 2006, \apjs, 165, 229

\bibitem[{{Fox} {et~al.}(2005){Fox}, {Wakker}, {Savage}, {Tripp}, {Sembach}, \&
  {Bland-Hawthorn}}]{2005ApJ...630..332F}
{Fox}, A.~J., {Wakker}, B.~P., {Savage}, B.~D., {et~al.} 2005, \apj, 630, 332

\bibitem[{{Gardiner} \& {Noguchi}(1996)}]{1996MNRAS.278..191G}
{Gardiner}, L.~T., \& {Noguchi}, M. 1996, \mnras, 278, 191

\bibitem[{{Haardt} \& {Madau}(2001)}]{2001cghr.confE..64H}
{Haardt}, F., \& {Madau}, P. 2001, in CGHR, ed. {D.~M.~Neumann \&
  J.~T.~V.~Tran}

\bibitem[{{Haffner}(2005)}]{2005ASPC..331...25H}
{Haffner}, L.~M. 2005, in ASP Conf. Ser., Vol. 331, Extra-Planar Gas, ed.
  {R.~Braun}, 25

\bibitem[{{Haffner} {et~al.}(1998){Haffner}, {Reynolds}, \&
  {Tufte}}]{1998ApJ...501L..83H}
{Haffner}, L.~M., {Reynolds}, R.~J., \& {Tufte}, S.~L. 1998, \apjl, 501, L83

\bibitem[{{Haffner} {et~al.}(1999){Haffner}, {Reynolds}, \&
  {Tufte}}]{1999ApJ...523..223H}
---. 1999, \apj, 523, 223

\bibitem[{{Haffner} {et~al.}(2001){Haffner}, {Reynolds}, \&
  {Tufte}}]{2001ApJ...556L..33H}
---. 2001, \apjl, 556, L33

\bibitem[{{Haffner} {et~al.}(2003){Haffner}, {Reynolds}, {Tufte}, {Madsen},
  {Jaehnig}, \& {Percival}}]{2003ApJS..149..405H}
{Haffner}, L.~M., {Reynolds}, R.~J., {Tufte}, S.~L., {et~al.} 2003, \apjs, 149,
  405

\bibitem[{{Hartmann, D.~\& Burton, W.~B.}(1997)}]{1997agnh.book.....H}
{Hartmann, D.~\& Burton, W.~B.}, ed. 1997, {Atlas of Galactic Neutral Hydrogen
  (Cambridge: Cambridge University Press)}

\bibitem[{{Hausen} {et~al.}(2002){Hausen}, {Reynolds}, {Haffner}, \&
  {Tufte}}]{2002ApJ...565.1060H}
{Hausen}, N.~R., {Reynolds}, R.~J., {Haffner}, L.~M., \& {Tufte}, S.~L. 2002,
  \apj, 565, 1060

\bibitem[{{Hill} {et~al.}(2008){Hill}, {Benjamin}, {Kowal}, {Reynolds},
  {Haffner}, \& {Lazarian}}]{2008ApJ...686..363H}
{Hill}, A.~S., {Benjamin}, R.~A., {Kowal}, G., {et~al.} 2008, \apj, 686, 363

\bibitem[{{Hill} {et~al.}(2009){Hill}, {Haffner}, \&
  {Reynolds}}]{2009ApJ...703.1832H}
{Hill}, A.~S., {Haffner}, L.~M., \& {Reynolds}, R.~J. 2009, \apj, 703, 1832

\bibitem[{{Kalberla} {et~al.}(2005){Kalberla}, {Burton}, {Hartmann}, {Arnal},
  {Bajaja}, {Morras}, \& {P{\"o}ppel}}]{2005A&A...440..775K}
{Kalberla}, P.~M.~W., {Burton}, W.~B., {Hartmann}, D., {et~al.} 2005, \aap,
  440, 775

\bibitem[{{Kunth} {et~al.}(1994){Kunth}, {Lequeux}, {Sargent}, \&
  {Viallefond}}]{1994A&A...282..709K}
{Kunth}, D., {Lequeux}, J., {Sargent}, W.~L.~W., \& {Viallefond}, F. 1994,
  \aap, 282, 709

\bibitem[{{Larson} {et~al.}(1980){Larson}, {Tinsley}, \&
  {Caldwell}}]{1980ApJ...237..692L}
{Larson}, R.~B., {Tinsley}, B.~M., \& {Caldwell}, C.~N. 1980, \apj, 237, 692

\bibitem[{Lehner \& Howk(2007)}]{2007MNRAS.377..687L}
Lehner, N., \& Howk, J.~C. 2007, \mnras, 377, 687

\bibitem[{{Lehner} \& {Howk}(2011)}]{2011Sci...334..955L}
{Lehner}, N., \& {Howk}, J.~C. 2011, Science, 334, 955

\bibitem[{{Lehner} {et~al.}(2012{\natexlab{a}}){Lehner}, {Howk}, {Thom}, {Fox},
  {Tumlinson}, {Tripp}, \& {Meiring}}]{2012MNRAS.424.2896L}
{Lehner}, N., {Howk}, J.~C., {Thom}, C., {et~al.} 2012{\natexlab{a}}, \mnras,
  424, 2896

\bibitem[{{Lehner} {et~al.}(2012{\natexlab{b}}){Lehner}, {Howk}, {Thom}, {Fox},
  {Tumlinson}, {Tripp}, \& {Meiring}}]{Lehner2012}
---. 2012{\natexlab{b}}, \mnras, submitted

\bibitem[{{Lin} \& {Lynden-Bell}(1977)}]{1977MNRAS.181...59L}
{Lin}, D.~N.~C., \& {Lynden-Bell}, D. 1977, \mnras, 181, 59

\bibitem[{{Lin} \& {Lynden-Bell}(1982)}]{1982MNRAS.198..707L}
---. 1982, \mnras, 198, 707

\bibitem[{{Lockman} {et~al.}(2008){Lockman}, {Benjamin}, {Heroux}, \&
  {Langston}}]{2008ApJ...679L..21L}
{Lockman}, F.~J., {Benjamin}, R.~A., {Heroux}, A.~J., \& {Langston}, G.~I.
  2008, \apjl, 679, L21

\bibitem[{{Madsen} {et~al.}(2006){Madsen}, {Reynolds}, \&
  {Haffner}}]{2006ApJ...652..401M}
{Madsen}, G.~J., {Reynolds}, R.~J., \& {Haffner}, L.~M. 2006, \apj, 652, 401

\bibitem[{{Martin}(1988)}]{1988ApJS...66..125M}
{Martin}, P.~G. 1988, \apjs, 66, 125

\bibitem[{{Meurer} {et~al.}(1985){Meurer}, {Bicknell}, \&
  {Gingold}}]{1985PASAu...6..195M}
{Meurer}, G.~R., {Bicknell}, G.~V., \& {Gingold}, R.~A. 1985, PASP, 6, 195

\bibitem[{{Moore} \& {Davis}(1994)}]{1994MNRAS.270..209M}
{Moore}, B., \& {Davis}, M. 1994, \mnras, 270, 209

\bibitem[{{Murai} \& {Fujimoto}(1980)}]{1980PASJ...32..581M}
{Murai}, T., \& {Fujimoto}, M. 1980, \pasj, 32, 581

\bibitem[{{Newberg} {et~al.}(2010){Newberg}, {Willett}, {Yanny}, \&
  {Xu}}]{2010ApJ...711...32N}
{Newberg}, H.~J., {Willett}, B.~A., {Yanny}, B., \& {Xu}, Y. 2010, \apj, 711,
  32

\bibitem[{{Oort}(1966)}]{1966BAN....18..421O}
{Oort}, J.~H. 1966, \bain, 18, 421

\bibitem[{{Oort}(1970)}]{1970A&A.....7..381O}
---. 1970, \aap, 7, 381

\bibitem[{{Pagel} \& {Patchett}(1975)}]{1975MNRAS.172...13P}
{Pagel}, B.~E.~J., \& {Patchett}, B.~E. 1975, \mnras, 172, 13

\bibitem[{{Putman} {et~al.}(2003){Putman}, {Bland-Hawthorn}, {Veilleux},
  {Gibson}, {Freeman}, \& {Maloney}}]{2003ApJ...597..948P}
{Putman}, M.~E., {Bland-Hawthorn}, J., {Veilleux}, S., {et~al.} 2003, \apj,
  597, 948

\bibitem[{{Raymond}(1979)}]{1979ApJS...39....1R}
{Raymond}, J.~C. 1979, \apjs, 39, 1

\bibitem[{{Reynolds}(1985)}]{1985ApJ...294..256R}
{Reynolds}, R.~J. 1985, \apj, 294, 256

\bibitem[{{Reynolds}(1991)}]{1991ApJ...372L..17R}
---. 1991, \apjl, 372, L17

\bibitem[{{Reynolds} {et~al.}(1999){Reynolds}, {Haffner}, \&
  {Tufte}}]{1999ASPC..168..149R}
{Reynolds}, R.~J., {Haffner}, L.~M., \& {Tufte}, S.~L. 1999, in ASP Conf. Ser.,
  Vol. 168, New Perspectives on the Interstellar Medium, ed. {A.~R.~Taylor,
  T.~L.~Landecker, \& G.~Joncas}, 149--+

\bibitem[{{Ryans} {et~al.}(1997){Ryans}, {Keenan}, {Sembach}, \&
  {Davies}}]{1997MNRAS.289..986R}
{Ryans}, R.~S.~I., {Keenan}, F.~P., {Sembach}, K.~R., \& {Davies}, R.~D. 1997,
  \mnras, 289, 986

\bibitem[{{Sancisi} {et~al.}(2008){Sancisi}, {Fraternali}, {Oosterloo}, \& {van
  der Hulst}}]{2008A&ARv..15..189S}
{Sancisi}, R., {Fraternali}, F., {Oosterloo}, T., \& {van der Hulst}, T. 2008,
  \aapr, 15, 189

\bibitem[{{Schmidt}(1963)}]{1963ApJ...137..758S}
{Schmidt}, M. 1963, \apj, 137, 758

\bibitem[{{Schwarz} {et~al.}(1995){Schwarz}, {Wakker}, \& {van
  Woerden}}]{1995A&A...302..364S}
{Schwarz}, U.~J., {Wakker}, B.~P., \& {van Woerden}, H. 1995, \aap, 302, 364

\bibitem[{{Sembach} {et~al.}(2000){Sembach}, {Howk}, {Ryans}, \&
  {Keenan}}]{2000ApJ...528..310S}
{Sembach}, K.~R., {Howk}, J.~C., {Ryans}, R.~S.~I., \& {Keenan}, F.~P. 2000,
  \apj, 528, 310

\bibitem[{{Sembach} {et~al.}(2003){Sembach}, {Wakker}, {Savage}, {Richter},
  {Meade}, {Shull}, {Jenkins}, {Sonneborn}, \& {Moos}}]{2003ApJS..146..165S}
{Sembach}, K.~R., {Wakker}, B.~P., {Savage}, B.~D., {et~al.} 2003, \apjs, 146,
  165

\bibitem[{{Shull} {et~al.}(2009){Shull}, {Jones}, {Danforth}, \&
  {Collins}}]{2009ApJ...699..754S}
{Shull}, J.~M., {Jones}, J.~R., {Danforth}, C.~W., \& {Collins}, J.~A. 2009,
  \apj, 699, 754

\bibitem[{{Spitzer}(1956)}]{1956ApJ...124...20S}
{Spitzer}, Jr., L. 1956, \apj, 124, 20

\bibitem[{{Thilker} {et~al.}(2005){Thilker}, {Braun}, \&
  {Westmeier}}]{2005ASPC..331..113T}
{Thilker}, D.~A., {Braun}, R., \& {Westmeier}, T. 2005, in ASP Conf. Ser., Vol.
  331, Extra-Planar Gas, ed. {R.~Braun}, 113

\bibitem[{{Tufte} {et~al.}(1998){Tufte}, {Reynolds}, \&
  {Haffner}}]{1998ApJ...504..773T}
{Tufte}, S.~L., {Reynolds}, R.~J., \& {Haffner}, L.~M. 1998, \apj, 504, 773

\bibitem[{{van den Bergh}(1962)}]{1962AJ.....67..486V}
{van den Bergh}, S. 1962, \aj, 67, 486

\bibitem[{{van Woerden} {et~al.}(1999){van Woerden}, {Schwarz}, {Peletier},
  {Wakker}, \& {Kalberla}}]{1999Natur.400..138V}
{van Woerden}, H., {Schwarz}, U.~J., {Peletier}, R.~F., {Wakker}, B.~P., \&
  {Kalberla}, P.~M.~W. 1999, \nat, 400, 138

\bibitem[{{van Woerden} \& {Wakker}(2004)}]{2004ASSL..312..195V}
{van Woerden}, H., \& {Wakker}, B.~P. 2004, in ASSL, Vol. 312, High Velocity
  Clouds, ed. {H.~van Woerden, B.~P.~Wakker, U.~J.~Schwarz, \& K.~S.~de Boer },
  195--226

\bibitem[{{Wakker} {et~al.}(1996){Wakker}, {Howk}, {Schwarz}, {van Woerden},
  {Beers}, {Wilhelm}, {Kalberla}, \& {Danly}}]{1996ApJ...473..834W}
{Wakker}, B., {Howk}, C., {Schwarz}, U., {et~al.} 1996, \apj, 473, 834

\bibitem[{{Wakker}(2001)}]{2001ApJS..136..463W}
{Wakker}, B.~P. 2001, \apjs, 136, 463

\bibitem[{{Wakker} {et~al.}(2001){Wakker}, {Kalberla}, {van Woerden}, {de
  Boer}, \& {Putman}}]{2001ApJS..136..537W}
{Wakker}, B.~P., {Kalberla}, P.~M.~W., {van Woerden}, H., {de Boer}, K.~S., \&
  {Putman}, M.~E. 2001, \apjs, 136, 537

\bibitem[{{Wakker} \& {Savage}(2009)}]{2009ApJS..182..378W}
{Wakker}, B.~P., \& {Savage}, B.~D. 2009, \apjs, 182, 378

\bibitem[{{Wakker} {et~al.}(2005){Wakker}, {Savage}, {Sembach}, {Richter}, \&
  {Fox}}]{2005ASPC..331...11W}
{Wakker}, B.~P., {Savage}, B.~D., {Sembach}, K.~R., {Richter}, P., \& {Fox},
  A.~J. 2005, in ASP Conf. Ser., Vol. 331, Extra-Planar Gas, ed. {R.~Braun}, 11

\bibitem[{{Wakker} {et~al.}(2008){Wakker}, {York}, {Wilhelm}, {Barentine},
  {Richter}, {Beers}, {Ivezi{\'c}}, \& {Howk}}]{2008ApJ...672..298W}
{Wakker}, B.~P., {York}, D.~G., {Wilhelm}, R., {et~al.} 2008, \apj, 672, 298

\bibitem[{{Wakker} {et~al.}(2003){Wakker}, {Savage}, {Sembach}, {Richter},
  {Meade}, {Jenkins}, {Shull}, {Ake}, {Blair}, {Dixon}, {Friedman}, {Green},
  {Green}, {Kruk}, {Moos}, {Murphy}, {Oegerle}, {Sahnow}, {Sonneborn},
  {Wilkinson}, \& {York}}]{2003ApJS..146....1W}
{Wakker}, B.~P., {Savage}, B.~D., {Sembach}, K.~R., {et~al.} 2003, \apjs, 146,
  1

\bibitem[{{Westmeier} {et~al.}(2005){Westmeier}, {Br{\"u}ns}, \&
  {Kerp}}]{2005ASPC..331..105W}
{Westmeier}, T., {Br{\"u}ns}, C., \& {Kerp}, J. 2005, in ASP Conf. Ser., Vol.
  331, Extra-Planar Gas, ed. {R.~Braun} ({San Francisco: ASP}), 105

\bibitem[{{Wood} {et~al.}(2010){Wood}, {Hill}, {Joung}, {Mac Low}, {Benjamin},
  {Haffner}, {Reynolds}, \& {Madsen}}]{2010ApJ...721.1397W}
{Wood}, K., {Hill}, A.~S., {Joung}, M.~R., {et~al.} 2010, \apj, 721, 1397

\end{thebibliography}

\clearpage

\begin{center}
\begin{deluxetable*}{lccccccccccccc}
\tabletypesize{\scriptsize}
 \tablewidth{500pt}
\tablecaption{Pointed intensities\label{table:intensity_time}}
 \tablehead{
   \multicolumn{3}{c}{} &
   \multicolumn{2}{c}{\ha} &
   \colhead{} &
   \multicolumn{2}{c}{[S{\sc~ii]} $\lambda6716$} &
   \colhead{} &
   \multicolumn{2}{c}{[N{\sc~ii]} $\lambda6584$} &
   \colhead{} &
   \multicolumn{2}{c}{[O{\sc~i]} $\lambda6300$} \\
   \cline{4-5} \cline{7-8} \cline{10-11} \cline{13-14}
   \colhead{} &
   \colhead{Position} &
   \colhead{\vhi} &  
   \colhead{I\tablenotemark{a}} &
   \colhead{t$_{\rm exp}$\tablenotemark{c}} &
   \colhead{} &
   \colhead{I\tablenotemark{a}} &
   \colhead{t$_{\rm exp}$\tablenotemark{c}} &
   \colhead{} &
   \colhead{I\tablenotemark{a}} &
   \colhead{t$_{\rm exp}$\tablenotemark{c}} &
   \colhead{} &   
   \colhead{I\tablenotemark{a}} &
   \colhead{t$_{\rm exp}$\tablenotemark{c}} \\
   \colhead{} &
   \colhead{$[\ell, b]$} &
   \colhead{[\kms]} &
   \colhead{[mR]} &
   \colhead{[min]} &
   \colhead{} &
   \colhead{[mR]} &
   \colhead{[min]} &
    \colhead{} &
   \colhead{[mR]} &
   \colhead{[min]} &   
   \colhead{} &
   \colhead{[mR]} &
   \colhead{[min]} 
   }
 \startdata
A0	 						& $134\fdg0, 25\fdg0$ 	& $-174.4\pm0.2$ 	& $<10$  					& 28		&& 				&		&& 				&		&& 			&	 \\	
AI 							& $139\fdg5, 29\fdg0$ 	& $-183.0\pm0.2$ 	& $<22$   					& 26	 	&& 				&		&& 				&		&& 			&	 \\
AII 							& $143\fdg5, 32\fdg5$ 	& $-142.4\pm0.2$	& $24\pm3\pm5$ 			& 26	 	&& 	 			&		&& 				&		&& 			&	 \\
AIII 							& $148\fdg0, 33\fdg7$ 	& $-154.6\pm0.1$ 	& $59\pm3\pm10$ 			& 26	  	&& 	 			&		&&  				&		&& 			&	 \\
AIII\tablenotemark{\textdagger}  	& $148\fdg5, 34\fdg5$ 	& $-153.0\pm0.1$ 	& $67\pm2\pm3$ 			& 36	 	&& $<26$			& 28		&& $26\pm2\pm9$	& 24		&& $<40$		& 6	 \\
AIV 							& $153\fdg6, 38\fdg2$ 	& $-173.9\pm0.1$ 	& $37\pm3\pm9$			& 26	 	&& 	 			&		&& 				&		&& 			& 	 \\
AIV\tablenotemark{\textdagger} 		& $153\fdg0, 38\fdg5$ 	& $-175.4\pm0.1$ 	& $50\pm2\pm14$ 			& 36	  	&& $16\pm6\pm11$	& 28		&& $<12$	  		& 24		&& $<91$		& 6	 \\
AV 							& $157\fdg0, 39\fdg6$ 	& $-159.9\pm0.2$ 	& $<17$   					& 24	  	&& 	 			&		&& 		 		&		&& 			&	 \\
AVI 							& $160\fdg3, 43\fdg3$ 	& $-157.3\pm0.2$ 	& $59\pm2\pm5$ 			& 24	  	&& 				&		&& 		 		&		&& 			&	 \\
Mrk 106 					 	& $161\fdg1, 42\fdg9$ 	& $-158.0\pm0.2$ 	& $39\pm3$\tablenotemark{b} 	&		&& $<36$			& 10		&& $33\pm3\pm5$ 	& 24		&& $<120$	& 6	 \\
Mrk 116 						& $165\fdg5, 44\fdg8$ 	& $-163.1\pm0.5$ 	& $32\pm6\pm13$			& 26		&& $25\pm6\pm16$	& 10		&& $<22$			& 24		&& $<125$	& 6	 \\
Mrk 25 						& $152\fdg7, 46\fdg8$ 	& --- 	       			& ---						& 24	  	&				& 		& &				& 		&&			&	 \\
PG 0822+645 					& $151\fdg6, 34\fdg6$ 	& $-168.4\pm0.2$ 	& $<29$\tablenotemark{b}  	&		&& 	 			& 10		&&  				& 24		&& $<88$		& 6	 \\
PG 0832+675 					& $147\fdg8, 35\fdg0$ 	& $-149.9\pm0.2$ 	& $25\pm3$\tablenotemark{b} 	&		&& $<31$			& 10		&& $<34$ 		& 24		&& $<139$	& 6	 \\
PG 0836+619 					& $154\fdg5, 36\fdg6$ 	& $-173.3\pm0.3$ 	& $<17$\tablenotemark{b}   	&		&& $<34$			& 10		&& $<28$			& 24		&& $<86$		& 6	 \\
$IRAS_{08339+6517}$ 			& $150\fdg5, 35\fdg6$ 	& $-156.3\pm0.5$ 	& $<28$\tablenotemark{b}   	&		& & $<61$ 		& 10		&& $<63$	 		& 24		&& $<125$	& 6	 \\
A$_{\rm low}$ 					& $137\fdg9, 27\fdg5$ 	& $-180.4\pm0.6$ 	& $<17$   					& 26		&				&		&& 		 		&		&&			&	 \\
UGC 4483 					& $145\fdg0, 34\fdg4$ 	& $-136.0\pm2.0$ 	& $<21$  					& 26		& 				&		&&   	 			&		&& 			&	 \\
A$_{\perp1}$					& $138\fdg3, 31\fdg0$ 	& $-157.0\pm2.0$ 	& $<15$  					& 26		&				&		&& 	 			&		&&			&	 \\
A$_{\perp2}$	 				& $139\fdg7, 30\fdg3$ 	& $-167.8\pm0.4$ 	& $<15$   					& 28		& 				&		&&   	 			&		&&			&	 \\
A$_{\perp3}$					& $141\fdg1, 29\fdg5$ 	& $-164.7\pm0.3$ 	& $<25$   					& 26		& 				&		&&   	 			&		&& 			&	 \\
A$_{\perp4}$	 				& $142\fdg5, 28\fdg8$ 	& $-171.8\pm0.4$ 	& $<27$   					& 26		& 				&		&&  	 			&		&& 			&	 \\
A$_{\perp5}$	 				& $144\fdg0, 28\fdg0$ 	& --- 		 		& $<17$   					& 26		& 				&		&&  	 			&		&&			&	 \\
Knot	 						& $159\fdg3, 35\fdg7$ 	& $-159.7\pm0.5$ 	& $<25$  					& 24		&  				&		&&  	 			&		&& 			&	 \\
UGC 4305 					& $144\fdg3, 32\fdg7$ 	& $-142.9\pm0.3$ 	& $<16$   					& 24		& 				&		&& 	 			&		&& 			&	 \\
A$_{26}$	 					& $136\fdg3, 26\fdg6$ 	& $-155.6\pm1.8$ 	& $<23$ 					& 24		& 				&		&&  	 			&		&& 			&	 \\
A$_{31}$	 					& $149\fdg0, 31\fdg0$ 	& $-175.4\pm0.4$ 	& $<16$   					& 26		&  				&		&&  	 			&		&& 			&	 \\
A$_{33}$	 					& $147\fdg0, 32\fdg2$ 	& $-147.5\pm0.5$ 	& $<32$   					& 26		& 				&		&&  	 			&		&& 			&	 \\
A$_{36}$						& $150\fdg0, 36\fdg0$ 	& $-148.8\pm0.3$ 	& $46\pm3\pm8$  			& 26		&				&		&&   	 			&		&& 			&	 \\
A$_{38}$	 					& $154\fdg0, 38\fdg5$ 	& $-175.5\pm0.1$ 	& $<26$   					& 26		&  				&		&& 	 			&		&& 			&	 \\
A$_{41}$	 					& $159\fdg0, 41\fdg0$ 	& $-155.9\pm0.2$ 	& $28\pm4\pm8$ 			& 24		& 				&		&& 	 	 		&		&& 			&	 \\
 \enddata
 \tablenotetext{a}{Non-extinction corrected intensities. The first set of errors represent statistical uncertainties; the second set denotes the uncertainty in measuring the intensity. These values are dominated by systematics associated with differences between the ons and offs (see Section \ref{section:pointed}); the mapped \ha\ observations presented in Figure \ref{figure:Map} do not suffer from the same effects as they are reduced using an atmospheric template.  }
 \tablenotetext{b}{Intensity from the mapped \ha\ data set.}
\tablenotetext{c}{Each individual exposure is 120 seconds long; the values listed represent the integrated exposure time.}
\tablenotetext{d}{Dates observed have been excluded from the arXiv version of this paper so that the table will fit.}
\end{deluxetable*}
\end{center}

\begin{deluxetable}{lcccccccccccccccc}
\tablecolumns{12}
\tabletypesize{\scriptsize}
\tablecaption{Neutral and Ionized Properties\label{table:mass}} 
\tablewidth{0pt}
\tablehead{
\multicolumn{1}{c}{} & \colhead{} & \multicolumn{3}{c}{Neutral Properties} &\colhead{}  & \multicolumn{2}{c}{Ionized Skin (n$_e=1/2n_0$)} & \colhead{} &\multicolumn{2}{c}{Ionized Skin (n$_e=n_0$)} & \colhead{} & \multicolumn{2}{c}{Mixed ($L_{\rm H^+}$=\lhi)}\\ 
\cline{3-5} \cline{7-8} \cline{10-11} \cline{13-14}
\colhead{Region\tablenotemark{a}}     	& \colhead{}	& \colhead{$M_{{\rm H}^0}$\tablenotemark{c}} 	& \colhead{$\log$~\lhi}		& \colhead{$\log\ \langle{n_0}\rangle$}   & \colhead{$\langle{EM}\rangle$\tablenotemark{d}} 		& \colhead{$\log\ L_{\rm H^+}$\tablenotemark{d}}	& \colhead{$M_{\rm {H^+}}$\tablenotemark{d}} & \colhead{} & \colhead{$\log\ L_{\rm H^+}$\tablenotemark{d}}	& \colhead{$M_{\rm {H^+}}$\tablenotemark{d}}  & \colhead{} 	& \colhead{$\log\ \langle n_e\rangle$\tablenotemark{d}}& \colhead{$M_{\rm {H^+}}$\tablenotemark{d}}	& \colhead{assumed $d$} \\	
\colhead{}     	& \colhead{}	& \colhead{[$10^5\ M_{\odot}$]} 	& \colhead{[kpc]}		& \colhead{[$\cm^{-3}$]}	& \colhead{[$10^{-3}\ \pc \cm^{-6}$]}  		&\colhead{[\kpc]} & \colhead{[10$^5\ M_{\odot}$]} & \colhead{} &\colhead{[\kpc]} & \colhead{[10$^5\ M_{\odot}$]}	& \colhead{} 	&\colhead{[$\cm^{-3}$]}	& \colhead{[$10^5\ M_{\odot}$]}  & \colhead{[\kpc]}}
\startdata	
\multicolumn{15}{c}{Distance to each region reflects a projected angle anchored by measured distances\tablenotemark{e}}  \\ \cline{1-15}		  		 
A0		&		& $0.5$	& $2.4$	& $-1.5$		& $136$	& 2.6 	& 0.7 	&& $2.0$		&0.3 && $-1.7$		& 0.5 & $6.1$	\\
AI		&		& $2.2$	& $2.7$	& $-1.6$		& $111$	& 2.7 	& 1.7 	&& $2.1$		&0.9 && $-1.8$		& 1.5 & $6.9$	\\
AII		&		& $1.1$	& $2.6$	& $-1.8$		& $74$	& 3.1 	& 2.1 	&& $2.5$		&1.1 && $-1.9$		& 1.2 & $7.5$	\\
AIII		&		& $2.0$	& $2.6$	& $-1.6$		& $97$	& 2.6 	& 1.4 	&& $2.0$		&0.7 && $-1.9$		& 1.3 & $8.0$\,\tablenotemark{e}\\
AIV		&		& $4.5$	& $2.8$	& $-1.8$		& $77$	& 2.9 	& 4.4 	&& $2.3$		&2.2 && $-2.0$		& 3.6 & $8.6$	\\
AV		&		& $1.3$	& $2.7$	& $-1.8$		& $97$	& 3.1		& 3.0 	&& $2.5$		&1.5 && $-1.9$		& 1.6 & $9.3$\,\tablenotemark{e}\\
AVI		&		& $2.4$	& $2.8$	& $-1.9$		& $123$	& 3.4 	& 7.4 	&& $2.8$		&3.7 && $-1.9$		& 3.4 & $9.9$\,\tablenotemark{e}\\
B      		&		& $1.6$	& $2.8$	& $-2.1$		& $81$	& 3.7 	& 9.4 	&& $3.1$		&4.7 && $-2.0$		& 3.1 & $9.9$\\				  	
\cline{1-15} \multicolumn{14}{c}{All regions assumed to have a distance of $9.0~\kpc$\,\tablenotemark{e}}  \\ \cline{1-15}
A0		&		& $1.1$	& $2.6$	& $-1.7$		& $136$	& 3.0 	& 2.2 	&& $2.4$		& 1.1 && $-2.9$		& 1.3 & $9.0$ \\
AI		&		& $3.7$	& $2.8$	& $-1.7$		& $111$	& 2.9 	& 3.8 	&& $2.3$		& 1.9 && $-2.5$	 	& 3.0 & $9.0$ \\
AII		&		& $1.6$	& $2.7$	& $-1.9$		& $74$	& 3.2 	& 3.7 	&& $2.6$	 	& 1.8 && $-2.7$		& 1.9 & $9.0$ \\
AIII		&		& $2.5$	& $2.7$	& $-1.6$		& $97$	& 2.7 	& 2.1 	 && $2.1$		& 1.0 && $-2.7$		& 1.8 & $9.0$\,\tablenotemark{e} \\
AIV	 	&		& $5.0$	& $2.8$	& $-1.8$		& $77$	& 3.0 	& 5.0 	&& $2.4$		& 2.5 && $-2.4$		& 4.0 & $9.0$ \\
AV		&		& $1.3$	& $2.7$	& $-1.8$		& $97$	& 3.1 	& 2.7 	&& $2.5$		& 1.3 && $-2.8$		& 1.5 & $9.0$\,\tablenotemark{e} \\
AVI		&		& $2.0$	& $2.7$	& $-1.8$		& $123$	& 3.3 	& 5.6 	&& $2.7$		& 2.8 && $-2.6$		& 2.7 & $9.0$\,\tablenotemark{e} \\
B		&		& $1.3$	& $2.8$	& $-2.1$		& $81$	& 3.6 	& 7.1 	&& $3.0$		& 3.5 &&  $-2.6$	& 2.4 & $9.0$ \\	
\cline{1-15} \multicolumn{14}{c}{All regions assumed to have a Galactic height $5.7~\kpc$ above the disk\,\tablenotemark{e,f}}  \\ \cline{1-15}
A0		&		& $2.6$	& $2.8$	& $-1.8$		& $136$	& 3.3 	&7.6	&& $2.7$		& 3.8 && $-1.9$	 	& 3.6 & $13.7$ \\
AI		&		& $6.4$	& $2.9$	& $-1.9$		& $111$	& 3.2 	&8.8	&& $2.6$		& 4.4 && $-2.0$	   	& 6.0 & $11.9$ \\
AII		&		& $2.3$	& $2.8$	& $-2.0$		& $74$	& 3.4 	&6.3	&& $2.8$	 	& 3.2 && $-2.0$	 	& 3.1 & $10.8$ \\
AIII		& 		& $3.1$	& $2.7$	& $-1.7$		& $97$	& 2.8  	&2.8	&& $2.2$		& 1.4 && $-1.9$	 	& 2.3 & $10.0$\,\tablenotemark{e} \\
AIV	 	&		& $5.2$	& $2.9$	& $-1.8$		& $77$	& 3.0 	&5.4	&& $2.2$		& 2.7 && $-1.8$	 	& 4.2 & $9.2$ \\
AV		&		& $1.2$	& $2.6$	& $-1.8$		& $97$	& 3.1 	&2.5	&& $2.4$		& 1.3 && $-1.8$	 	& 1.4 & $8.8$\,\tablenotemark{e} \\
AVI		&		& $1.7$	& $2.7$	& $-1.8$		& $123$	& 3.3 	&4.5	&& $2.7$		& 2.3 && $-1.8$	 	& 2.2 & $8.4$\,\tablenotemark{e} \\
B		&		& $1.3$	& $2.8$	& $-2.1$		& $81$	& 3.7 	&3.8	&& $3.1$		& 3.8 && $-2.1$ 	& 2.5 & $9.2$ \\	
\cline{1-15} \multicolumn{14}{c}{Distances from matching the MW and EGB ionizing flux with the $\Phi_{\rm LC}(\mha)$ \tablenotemark{e}}  \\ \cline{1-15}	
A0		&		& $0.6$	& $2.6$	& $-1.5$		& $136$	& 2.7 	& 0.8 	&& $2.1$		& 0.4 && $-1.8$	 	& 0.5 & $6.4$ \\
AI		&		& $2.2$	& $2.8$	& $-1.6$		& $111$	& 2.7 	& 2.2 	&& $2.1$		& 0.9 && $-1.9$	   	& 1.5 & $6.9$ \\
AII		&		& $2.0$	& $2.7$	& $-2.0$		& $74$	& 3.3 	& 5.0 	&& $2.7$	 	& 2.5 && $-2.0$	 	& 2.5 & $10.0$ \\
AIII		&		& $2.1$	& $2.7$	& $-1.6$		& $97$	& 2.7 	& 1.6 	&& $2.1$		& 0.8 && $-1.9$	 	& 1.5 & $8.3$\,\tablenotemark{e} \\
AIV		&		& $4.9$	& $2.8$	& $-1.8$		& $77$	& 3.0 	& 4.9 	&& $2.4$		& 2.4 && $-2.0$	 	& 3.9 & $8.9$ \\
AV		&		& $1.1$	& $2.7$	& $-1.8$		& $97$	& 3.1 	& 2.3 	&& $2.5$		& 1.1 && $-1.9$	 	& 1.3 & $8.5$\,\tablenotemark{e} \\
AVI		&		& $1.5$	& $2.7$	& $-1.8$		& $123$	& 3.2 	& 3.5 	&& $2.6$		& 1.7 && $-1.8$	 	& 1.8 & $7.7$\,\tablenotemark{e} \\
B		&		& $1.2$	& $2.8$	& $-2.1$		& $81$	& 3.6 	& 6.6 	&& $3.0$		& 3.3 &&  $-2.0$	& 2.3 & $8.8$ 	
\enddata
\tablenotetext{a}{These regions are defined in Table \ref{table:region} and shown graphically in Figure \ref{figure:Map}.} 
\tablenotetext{b}{The average uncertainty for the foreground H{\sc~i} is  $7.0\times10^{16}$~cm$^{-2}$ and $5.8\times10^{16}$~cm$^{-2}$ for Complex~A. The  The $\log\ \langle$\nhi$\rangle$ values for the foreground and internal cloud emission has been excluded from the arXiv version of this paper so that the table will fit.} 
\tablenotetext{c}{${M}_{{\rm H}^0}=1.4\ $\nhi\ $\Omega\ d^2$}
\tablenotetext{d}{Assumes an electron temperature of $10^4\rm{K}$.} 
\tablenotetext{e}{From absorption-line studies, the distance to core AIII is greater than 8.1~kpc, the distance to core AV is greater than $4.0\pm1.0$~kpc, and the distance to is  core AVI is less than $9.9\pm1.0$~kpc \citep{1996ApJ...473..834W, 1997MNRAS.289..986R, 1999Natur.400..138V, 2003ApJS..146....1W}.    }
\tablenotetext{f}{The constant height above the Galactic plane at $z=5.7~\kpc$ is defined by placing the midpoint of cores AIII and AVI at a distance of 9.0~\kpc.} 
\end{deluxetable}

\end{document}